\documentclass[a4paper,11pt]{article}
\pdfoutput=1
\usepackage{color}
\usepackage{epsfig}
\usepackage[caption = false]{subfig}
\usepackage{amsmath}
\usepackage{amssymb}
\usepackage{graphicx}
\usepackage{jheppub} % for details on the use of the package, please
\usepackage[T1]{fontenc}

\newcommand{\beq}{\begin{equation}}
\newcommand{\eeq}{\end{equation}}

\newcommand{\ou}{%
  \mathrel{%
    \vcenter{\offinterlineskip
      \ialign{##\cr$<$\cr\noalign{\kern-1.5pt}$>$\cr}%
    }%
  }%
}

\newcommand{\uo}{%
  \mathrel{%
    \vcenter{\offinterlineskip
      \ialign{##\cr$>$\cr\noalign{\kern-1.5pt}$<$\cr}%
    }%
  }%
}

\def\Xint#1{\mathchoice
   {\XXint\displaystyle\textstyle{#1}}%
   {\XXint\textstyle\scriptstyle{#1}}%
   {\XXint\scriptstyle\scriptscriptstyle{#1}}%
   {\XXint\scriptscriptstyle\scriptscriptstyle{#1}}%
   \!\int}
\def\XXint#1#2#3{{\setbox0=\hbox{$#1{#2#3}{\int}$}
     \vcenter{\hbox{$#2#3$}}\kern-.5\wd0}}

\def\dashint{\Xint-}

\def \p {\phi}

\title{\boldmath Continuum limit of fishnet graphs and AdS sigma model}

\author{Benjamin Basso and De-liang Zhong}

\affiliation{Laboratoire de Physique Th\'eorique de l'\'Ecole Normale Sup\'erieure, CNRS,\\
Universit\'e PSL, Sorbonne Universit\'es, Universit\'e Pierre et Marie Curie,\\
24 rue Lhomond, 75005 Paris, France}

\abstract{We consider the continuum limit of 4d planar fishnet diagrams using integrable spin chain methods borrowed from the $\mathcal{N}=4$ Super-Yang-Mills theory. These techniques give us control on the scaling dimensions of single-trace operators for all values of the coupling constant in the fishnet theory. We use them to study the thermodynamical limit of the BMN operator corresponding to the spin chain ferromagnetic vacuum. We find that its scaling dimension exhibits a critical behaviour when the coupling constant approaches Zamolodchikov's critical coupling. Analysis close to that point suggests that the continuum limit of the fishnet graphs is controlled by the two-dimensional $AdS_{5}$ non-linear sigma model. More generally, we present evidence that the fishnet diagrams define an integrable lattice regularization of the $AdS_{5}$ model. A system of massless TBA equations is derived for the tachyon energy by dualizing the TBA equations of the weakly coupled planar $\mathcal{N}=4$ SYM theory.}

\begin{document}

\maketitle
\flushbottom

\renewcommand{\thefootnote}{\arabic{footnote}}

\section{Introduction}\label{Sec1}

String sigma models are believed to provide a general solution for the sum over planar diagrams in gauge theories \cite{tHooft:1973alw}. The AdS-CFT correspondence \cite{Aharony:1999ti} shed light on this old idea and suggested new embodiments for conformally invariant gauge theories. The most famous example is the 4d $\mathcal{N}=4$ SYM theory which is conjectured to be dual to string theory in AdS \cite{Maldacena:1997re}. This supersymmetric gauge theory is also special in that it is believed to be integrable at large $N$ \cite{Beisert:2010jr}. The latter property gives us a handle on the AdS-CFT dictionary, enabling the development of new techniques for carrying out the large $N$ re-summation of the field theory diagrams, at both planar \cite{Beisert:2010jr,Basso:2013vsa,Basso:2015zoa,Fleury:2016ykk,Eden:2016xvg} and non-planar level \cite{Bargheer:2017nne,Eden:2017ozn,Ben-Israel:2018ckc}. Furthermore, these methods allow us to explore a larger chunk of the correspondence between planar diagrams and sigma models by means of partial or twisted re-summations, which are naturally associated with some integrable deformations of $\mathcal{N}=4$ SYM.

\begin{figure}
\begin{center}
\includegraphics[scale=0.6]{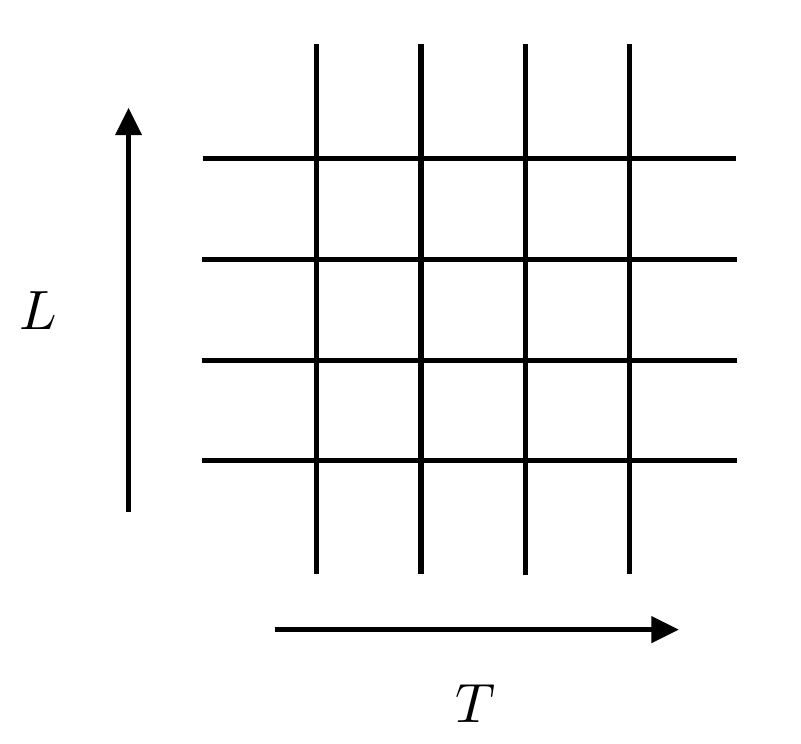}
\end{center}
%\vspace{-8.8cm}
\caption{A planar fishnet diagram of order $L\times T$. Every intersection point $x$ is integrated over spacetime and every connecting line stands for a massless scalar propagator $\sim 1/(x-y)^2$. In the fishnet theory, the horizontal and vertical lines correspond to trajectories of the $\phi_{1}$ and $\phi_{2}$ particles, respectively, and the diagram is weighted by $L\times T$ powers of the coupling $g^{2}$.}\label{fishnets} 
\end{figure}

In this paper we consider such an integrable daughter of $\mathcal{N}=4$ SYM, which comes with no gauge fields nor any clear-cut stringy interpretation. The theory to be studied was introduced recently by Gurdogan and Kazakov \cite{Gurdogan:2015csr} and is known as the fishnet theory, see also \cite{Zamolodchikov:1980mb} for earlier work and \cite{Caetano:2016ydc, Grabner:2017pgm,Kazakov:2018qbr,Kazakov:2018ugh} for further developments. It consists of two complex matrix scalar fields interacting by means of a single quartic coupling%
\footnote{A proper definition requires introducing double-trace couplings, which for the sake of conformality must be tuned to their critical values \cite{Grabner:2017pgm,Sieg:2016vap}, see also \cite{Pomoni:2008de}. We will not need to worry about them here.}
\beq\label{Lag}
\mathcal{L} = N\,\textrm{tr}\,(\partial_{\mu}\phi_{1}\partial_{\mu}\phi_{1}^{*}+\partial_{\mu}\phi_{2}\partial_{\mu}\phi_{2}^{*}+ (4\pi g)^2\, \phi_{1}\phi_{2}\phi^{*}_{1}\phi^{*}_{2})\, ,
\eeq
with the trace taken over the $N\times N$ matrix indices, which here are just flavour indices. It can be viewed as a truncation of weakly coupled $\mathcal{N}=4$ SYM in which gluons and gauginos are forcefully decoupled and only two of the three complex scalar fields are retained. The proper procedure goes through the extremal twisting \cite{Gurdogan:2015csr} of the $\gamma$-deformed SYM theory \cite{Leigh:1995ep,Lunin:2005jy,Frolov:2005dj,Beisert:2005if} which involves sending the YM coupling to zero and the deformation parameter $\gamma$ to $i\infty$, while keeping the suitably rescaled coupling $g^2$ fixed. Owing to this ``embedding'', the theory is expected to be conformally invariant and integrable for any $g^2$  \cite{Gromov:2017cja,Grabner:2017pgm}, at least in the planar regime. In fact, the integrability of the fishnet vertex was recognized by Zamolodchikov more than 40 years ago \cite{Zamolodchikov:1980mb}.

One appealing feature of the fishnet theory is that it produces many fewer graphs than $\mathcal{N}=4$ SYM. Quite often only a single graph contributes at a given order in perturbation theory, in the planar limit. The price to pay for this massive cut is the loss of unitarity, as the strict ordering of the fields in the potential clashes with the reality of the action. Another drawback is that the duality with string theory is uncertain, for the AdS radius is naively small. Still, the fishnet theory proves to be a remarkable testing ground for integrability, which, in turn, sheds light on families of conformal Feynman integrals \cite{Chicherin:2017frs,Chicherin:2017cns} and suggests new ways of evaluating them \cite{Gurdogan:2015csr,Caetano:2016ydc,Gromov:2017cja,Basso:2017jwq,Grabner:2017pgm,Kazakov:2018qbr}.

The planar diagrams of the theory, the fishnet graphs, are special in that they all look locally like the $L\times T$ square lattice shown in figure \ref{fishnets}. Accordingly, every diagram can be viewed as a partition function $Z_{L, T}$ for a 2d vertex model \cite{Zamolodchikov:1980mb}, with the bulk spacetime points acting as classical ``spins'', the propagators as nearest neighbour couplings and with the graph's external lines setting up the boundary conditions. Different observables of the planar fishnet theory correspond to different boundary conditions and all the graphs obeying the same boundary conditions are summed over.

An important observation concerning the large order behaviour of the fishnet diagrams was made by Zamolodchikov \cite{Zamolodchikov:1980mb}, see also \cite{Bazhanov:2016ajm} for a recent discussion, who computed, using integrable vertex model techniques, the free energy per site in the thermodynamical limit
\beq\label{Z-th}
\log{Z}_{L, T} \sim -LT\log{g^2_{cr}}\, ,
\eeq
for graphs subject to periodic boundary conditions, and found that
\beq\label{cr}
g_{cr} = \frac{\Gamma(\tfrac{3}{4})}{\sqrt{\pi}\Gamma(\tfrac{5}{4})} \simeq 0.76....
\eeq
This constant determines a critical coupling for thermodynamically large observables in the fishnet theory and one might expect, in analogy with matrix models \cite{Klebanov:1991qa,DiFrancesco:1993cyw,Ginsparg:1993is,Nakayama:2004vk}, that a ``dual'' continuum description is taking over at that point.%
\footnote{Note that the critical coupling does not refer to a point at which the 4d theory is becoming critical; the fishnet theory is conformal for any $g^2$. It is a point at which the planar diagrams become dense.}

In this paper we examine the thermodynamical limit of fishnet graphs using integrable methods borrowed from the $\mathcal{N}=4$ SYM theory and argue that the continuum description is given by the 2d (bosonic) $AdS_{5}$ sigma model.

\begin{figure}
\begin{center}
\includegraphics[scale=0.5]{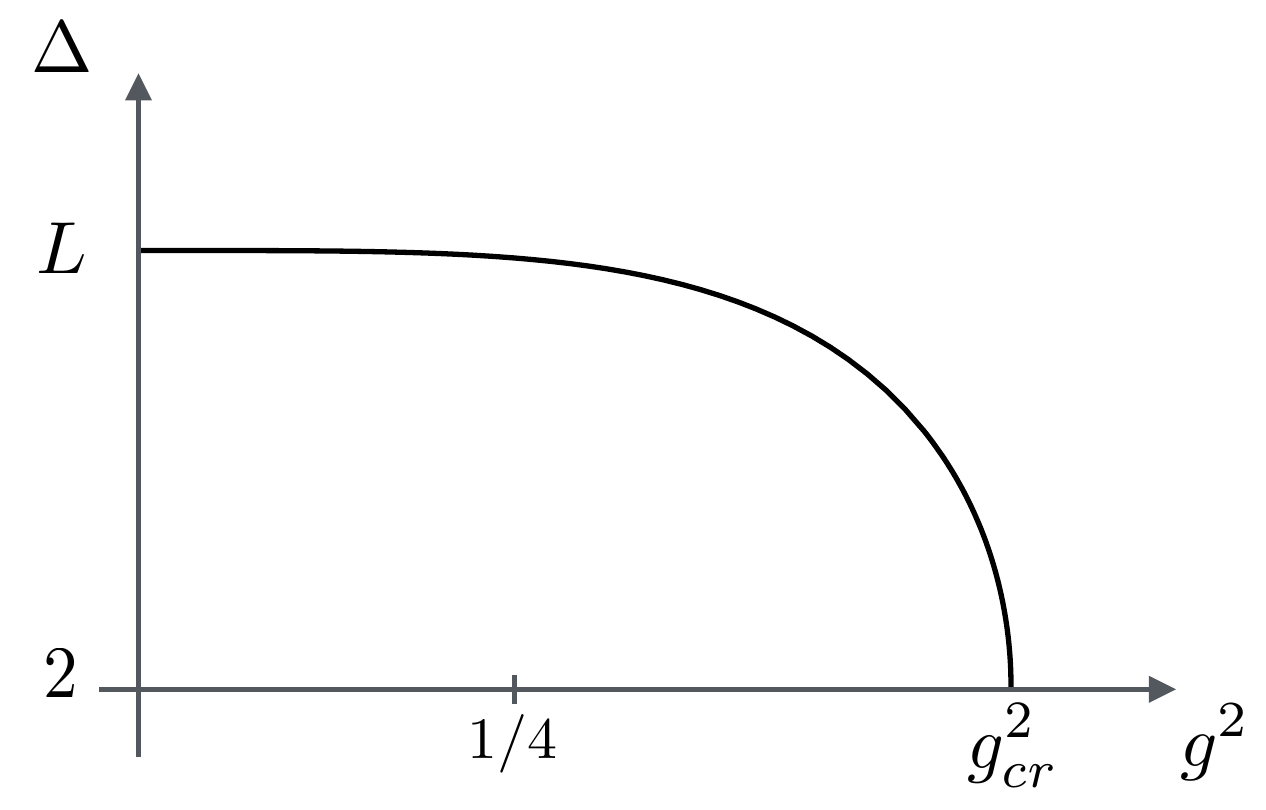}
\end{center}
%\vspace{-7.5cm}
\caption{Schematic plot of the scaling dimension $\Delta$ of the BMN operator $\textrm{tr}\, \phi_{1}^{L}$ as function of the coupling constant $g^2$, at large $L$. The scaling dimension is nearly independent of the coupling constant, $\Delta = L$, until $g = 1/2$ where it starts developing a non-trivial thermodynamical scaling, $\Delta \sim L f(g)$. At $g = g_{cr}$, the scaling function $f(g)$ vanishes and its derivative is infinite. The physics close to this point is controlled by the 2d low energy dynamics of a large fishnet graph. The overall shape of the thermodynamical curve agrees with the findings of \cite{Gromov:2017cja} for $L=3$. The value of the scaling dimension at the branch point appears to be $\Delta = 2$ at both small and large $L$.}\label{dimension}
\end{figure}

Our discussion will center around the scaling dimension $\Delta$ of the BMN operator $\textrm{tr}\, \phi_{1}^L$ which maps to the ground-state energy of a ferromagnetic non-compact spin chain \cite{Gromov:2017cja}. Integrability will allow us to study the thermodynamical limit of this energy for generic coupling $g$ by means of a linear integral equation. It will confirm the existence of a non-trivial thermodynamical scaling $\Delta \sim L f(g)$, for sufficiently ``strong'' coupling, and the emergence of a critical behaviour $\sim \sqrt{g_{cr}-g}$ close to the critical point, in line with the results of \cite{Gromov:2017cja,Grabner:2017pgm} for $L=2,3$.%
\footnote{The location of the branch point $g_{cr}(L)$ is function of the length $L$; in particular \cite{Grabner:2017pgm}, $g_{cr}(L=2) = 0$.} A sketch of the thermodynamical behaviour of the scaling dimension is shown in figure \ref{dimension}.

Dualizing our equation, by means of a particle-hole transformation, will reveal the nature of the critical point and suggest the interpretation of the BMN operator as describing the ``tachyon'' ground-state of the AdS sigma model,
\beq\label{corr}
\textrm{tr}\, \phi_{1}^L \qquad  \leftrightarrow \qquad V_{\Delta} \sim e^{-i\Delta t}\, ,
\eeq
labelled by the global time energy $\Delta$ of the BMN operator and implicitly by the size $L$ of the worldsheet. Though we will actually never cross the line where the AdS mass squared turns negative, we will stick to the name of tachyon for the dual object.%
\footnote{The tachyonic domain maps to $g > g_{cr}(L)$, where the scaling dimension has an imaginary part, $\Delta = 2+i\nu$; see \cite{Gromov:2017cja,Grabner:2017pgm,Pomoni:2008de} for discussions.}

The correspondence (\ref{corr}) is best summarized by the formula
\beq\label{log-g}
\log{g^{2L}} = \log{g_{cr}^{2L}} + E_{2d}(\Delta, L)\, ,
\eeq
which relates the 4d coupling $g^2$ to the sigma model energy $E_{2d}(\Delta, L)$ and shows that the vicinity of the critical point maps to low energies on the worldsheet. In this regime the $AdS_{5}$ sigma model is weakly coupled and we will be able to verify our claims directly.

We shall also test the correspondence at the level of the $1/L$ corrections and obtain a system of TBA equations for the tachyon ground state, valid in principle for any $L$ and $\Delta$. Its form will support the more general conjecture that the fishnet diagrams define an integrable lattice regularization of the $AdS_{5}$ sigma model. 

The plan is as follows. In Section \ref{Sect2} we introduce the integrability set up and derive the integral equation for the large $L$ limit of the scaling dimension $\Delta$. With its help we reproduce Zamolodchikov's prediction for the critical coupling. In Section \ref{Sect3} we obtain the dual system of equations, give arguments for its interpretation as describing the tachyon in the AdS sigma model and carry out some perturbative tests. In Section \ref{Sect4} we dualize the full system of TBA equations and compute the IR central charge. We conclude in Section \ref{Sect5}. The appendices contain a detailed analysis of the dual linear equation, a discussion of spinning operators and a brief study of the thermodynamical limit of 3d triangular fishnet diagrams.

\section{Thermodynamics of fishnet graphs}\label{Sect2}

We start with a light review of the integrability methods for computing the scaling dimension of interest, emphasizing the correspondence with the Feynman diagrams. Readers familiar with these techniques may jump directly to Subsections \ref{Sect2.2} and \ref{Sect2.3} where we restrict our attention to the thermodynamical limit and re-derive Zamolodchikov's critical coupling.

\subsection{Grand canonical ensemble}

In the planar $\mathcal{N}=4$ SYM integrable framework the BMN operator $\textrm{tr}\, \phi_{1}^{L}$ is identified with the ferromagnetic ground state of a periodic spin chain of length $L$ and its scaling dimension $\Delta$ maps to the spin chain vacuum energy. The same can be said in the fishnet theory. The main difference is that in the fishnet theory the vacuum is not protected and its energy $\Delta = \Delta_{L}(g)$ is a very complicated function of the length $L$ and coupling $g$. It was studied extensively in \cite{Gromov:2017cja,Grabner:2017pgm} for small values of $L$ and perturbatively at weak coupling for any $L$ in \cite{Gurdogan:2015csr}.

\begin{figure}
\begin{center}
\includegraphics[scale=0.4]{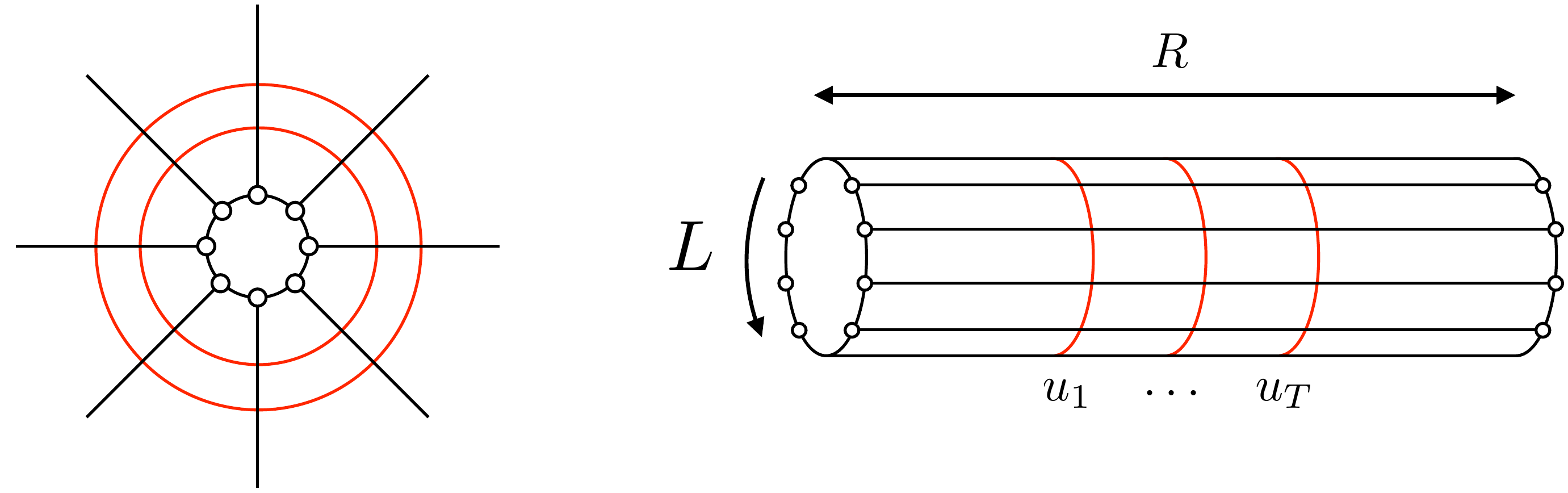}
\end{center}
%\vspace{-7cm}
\caption{On the left panel we show a two-wheel diagram contributing to the anomalous dimension of the BMN operator $\textrm{tr}\, \phi_{1}^L$, represented in the middle by a spin chain with $L$ sites. The operator is local but magnified here for the sake of illustration. Each wheel corresponds to the trajectory of a $\phi_{2}$ particle. On the right panel a wheel diagram is wrapped on a cylinder of size $R$ along the global time direction and length $L$ along the spin chain one. Taking $R\rightarrow \infty$ projects on the ground state at the boundary. Each vertical wheel gives rise to a tower of mirror magnons which are integrated over. Each magnon is labelled by a rapidity $u$, or momentum $p=2u$ along the time direction, and a Lorentz index $a =1, 2, ...$, which comes from the partial wave decomposition of the scalar field $\phi_{2}$ on $S^{3}$.}\label{cylinder0}
\end{figure}

For any length $L > 2$, otherwise see \cite{Grabner:2017pgm}, the diagrams contributing to the scaling dimension in the planar limit are the wheel diagrams \cite{Gurdogan:2015csr} where virtual $\phi_{2}$ particles loop around the operator $\textrm{tr}\, \phi_{1}^{L}$ as shown in figure \ref{cylinder0}. A general wheel diagram is obtained by pinching the end points of the $L$ horizontal lines in figure \ref{fishnets} and periodically identifying the vertical ones. Obviously, each wheel costs $L$ powers of the coupling $g^2$ and, consequently, the scaling dimension admits an expansion in powers of $g^{2L}$ at weak coupling, $\Delta = L+O(g^{2L})$, referred to as the wheel expansion in the following.

In the spin chain picture, the wheels map to long-range interactions mediated by virtual magnons travelling around the chain \cite{Ambjorn:2005wa,Bajnok:2008bm,Janik:2010kd}, also known as wrapping or L\"uscher corrections. The magnons circulating around the chain are not the familiar ones parameterizing spin waves on top of the ferromagnetic vacuum~\cite{Minahan:2002ve}. Instead they live in the orthogonal, so called mirror, kinematics where the time $t$ is interpreted as a space direction and the spin-chain length $L$ as an inverse temperature, see figure \ref{cylinder0}. The mirror picture, which in the general context is inherited from the dual string worldsheet theory, see \cite{Bajnok:2010ke,Arutyunov:2014cra} for further discussion, can be motivated by considering the problem on the 4d euclidean cylinder $\mathbb{R}\times S^3$, with $S^3$ the unit 3-sphere surrounding the operator and with the global time corresponding to the spin chain time $t$. The relevant $1+1$ dimensional picture is obtained by dropping the 3-sphere, which becomes internal, and replacing it by the spin-chain circle, $\mathbb{R}\times S^{3} \rightarrow \mathbb{R}\times S_{L}$.%
\footnote{$\mathbb{Z}_{L}$ would be more appropriate, but the difference is immaterial here.} 
The partition function on this geometry returns the ground state energy, in the large volume limit $R\rightarrow \infty$,
\beq\label{thermal}
\mathcal{Z}_{L, R}(g) = e^{-\Delta_{L}(g) R} + ...\, ,
\eeq
where the dots stand for heavier single-trace operators, with $\phi_{1}$ charge $L$, and where $R$ is the length of the cylinder along the $t$ direction.  The wheel expansion of the partition function, and thus of $\Delta$,
\beq
\mathcal{Z}_{L, R}(g) = \sum_{T\geqslant 0} g^{2L T}Z_{L, T}(R)\, ,
\eeq
is more naturally interpreted as the decomposition in the orthogonal, open string, channel, with the sum running over a complete basis of states along the $t$ direction, labelled by the total number of mirror magnons $T$. To be more precise, each wheel gives rise to a semi-infinite family of mirror magnons, stemming from the partial wave decomposition of the scalar field $\phi_{2}$ on $S^3$. Suppressing the Lorentz indices, each mirror magnon $\sim \partial^{a-1}\phi_{2}$ is tagged with an integer $a = 1, 2, ...$, for its Lorentz spins $(\tfrac{1}{2}(a-1), \tfrac{1}{2}(a-1))$ and, after smearing over the $t$ direction, with a momentum $p = 2u$, with $u\in \mathbb{R}$ the so called Bethe rapidity. The micro-canonical contribution $Z_{L, T}(R)$ follows then from integration and summation over the $T$ magnon phase space, with proper thermodynamical weights and measures; see \cite{Kostov:2018ckg} for a recent discussion of the relation between the micro-canonical partition functions and the canonical one.

The Thermodynamical Bethe Ansatz allows one to take advantage of the factorized scattering between mirror magnons and compute the bulk free energy (\ref{thermal}) to all orders in the wheel expansion and for any temperature $1/L$, see \cite{Ahn:2011xq,Bajnok:2010ke,vanTongeren:2016hhc} and references therein. The latter free energy takes the usual form
\beq\label{delta}
\Delta = L - \sum_{a\geqslant 1}\int_{-\infty}^{\infty}\frac{du}{2\pi} p'(u)\log{(1+Y_{a}(u))}\, ,
\eeq
with $p' =2$ in the case at hand and with the Y functions describing the thermal distribution of the energy among magnons. In the weak coupling regime $g^2 \ll 1$, the gas is rarefied and the TBA equations can be expanded around the Fermi-Dirac distribution, characterized by the Boltzmann weights $\textbf{Y}_{a}$,
\beq\label{bfYa}
Y_{a} \xrightarrow{g^2 \ll 1} \textbf{Y}_{a} = a^2g^{2L}e^{-L\epsilon_{a}(u)} \ll 1\, ,
\eeq
where $a^2$ is the dimension of the $a$-th Lorentz representation and where $\epsilon_{a}(u)$ is the mechanical energy of the associated magnon,
\beq\label{eps-a}
\epsilon_{a}(u) = \log{(u^2+a^2/4)}\, .
\eeq
 The departure from the free distribution is controlled by the interaction among magnons. To leading order, it takes the rather universal form
\beq\label{logY}
\log{Y_{a}(u)} = \log{\textbf{Y}_{a}(u)} + \sum_{b\geqslant 1}\int_{-\infty}^{\infty} \frac{dv}{2\pi} \textbf{Y}_{b}(v)(\frac{\partial}{i\partial v}\log{S_{b, a}(v, u)} + \ldots) + O(\textbf{Y}_{b}\textbf{Y}_{c})\, ,
\eeq
where $S_{a, b}$ is the dynamical factor of the magnon S-matrix, which specifies the model and reads
\beq\label{Sab}
S_{a, b}(u, v) = -\frac{u-v-i\tfrac{a+b}{2}}{u-v+i\tfrac{a+b}{2}}\prod_{k=0,1}\frac{\Gamma(k+\tfrac{a}{2}-iu)\Gamma(k+\tfrac{b}{2}+iv)\Gamma(k+\frac{a-b}{2}+iu-iv)}{\Gamma(k+\tfrac{a}{2}+iu)\Gamma(k+\tfrac{b}{2}-iv)\Gamma(k+\frac{a-b}{2}-iu+iv)}\, ,
\eeq
with $\Gamma(z)$ the Euler Gamma function. The dots in (\ref{logY}) capture the effect of the $O(4)$ degrees of freedom, or scattering of Lorentz indices, and are simply absent for the lightest magnons ($a=1$ or $b=1$), which are Lorentz singlets. (This isotopic component of the scattering is controlled by a rational $R$ matrix, which is function of the difference of rapidities and is coupling independent.)

Formula (\ref{logY}) is a particular case of the NLO L\"uscher formula for the ground state of the twisted $\mathcal{N}=4$ SYM spin chain \cite{Ahn:2011xq}, obtained by truncating the $\mathcal{N}=4$ SYM magnon super-multiplet to its $\phi_{2}$ component and by sending the coupling constant of the gauge theory to zero in all spectral and scattering data. Once plugged into (\ref{delta}), it yields the wheel expansion of the scaling dimension with the obvious map ``$\textbf{Y} = $ wheel'', such that the first term is the free wheel, the next one the double wheel, etc. It was used for comparison with the direct integration of the Feynman integrals in \cite{Gurdogan:2015csr}.

All the information for moving to higher orders is in principle contained in the TBA equations, which are pretty simple
\beq\label{TBA-1}
\log{Y_{a}(u)} = L\log{g^2}-L\epsilon_{a}(u) + \sum_{b\geqslant 1}\int_{-\infty}^{\infty} \frac{dv}{2\pi}\frac{\partial}{i\partial v}\log{S_{b, a}(v, u)}\log{(1+Y_{b}(v))} + \ldots.
\eeq
if not for the dots, which accommodate for the matrix degrees of freedom and cannot be spelled out without introducing an auxiliary set of variables. They will be given in their full form in Section \ref{Sect4}. Nonetheless, the general term in the wheel expansion (\ref{logY}) has not been worked out explicitly. Also, the integration over the magnons' phase space is nearly impossible in general. A more powerful analytical treatment relies on the Baxter equation \cite{Gromov:2017cja}, which relates to the (twisted) $\mathcal{N}=4$ SYM Quantum Spectral Curve \cite{Gromov:2013pga,Kazakov:2015efa} and enables higher loop computations of the scaling dimension at finite length $L$. We will not need so much improvement in this paper however, since in the thermodynamical limit $L\rightarrow \infty$ the TBA equations (\ref{TBA-1}) simplify drastically.

A distinguished feature of the fishnet TBA equations is that all the dependence on the coupling constant $g^2$ comes along with the energy, like in (\ref{TBA-1}) or equivalently (\ref{bfYa}). Put differently, the 4d coupling of the fishnet theory is a fugacity for the number of mirror magnons and we can think of the scaling dimension $\Delta$ as defining the free energy of a grand canonical ensemble at chemical potential $h = \log{g^2}$. This parallels the fact that the mirror magnons are in one-to-one correspondence with the wheels. Hence, one can easily probe fishnet graphs of arbitrarily large orders and obtain information about their continuum limit by playing with $h$.

\subsection{Thermodynamical limit}\label{Sect2.2}

\begin{figure}
\begin{center}
\includegraphics[scale=0.4]{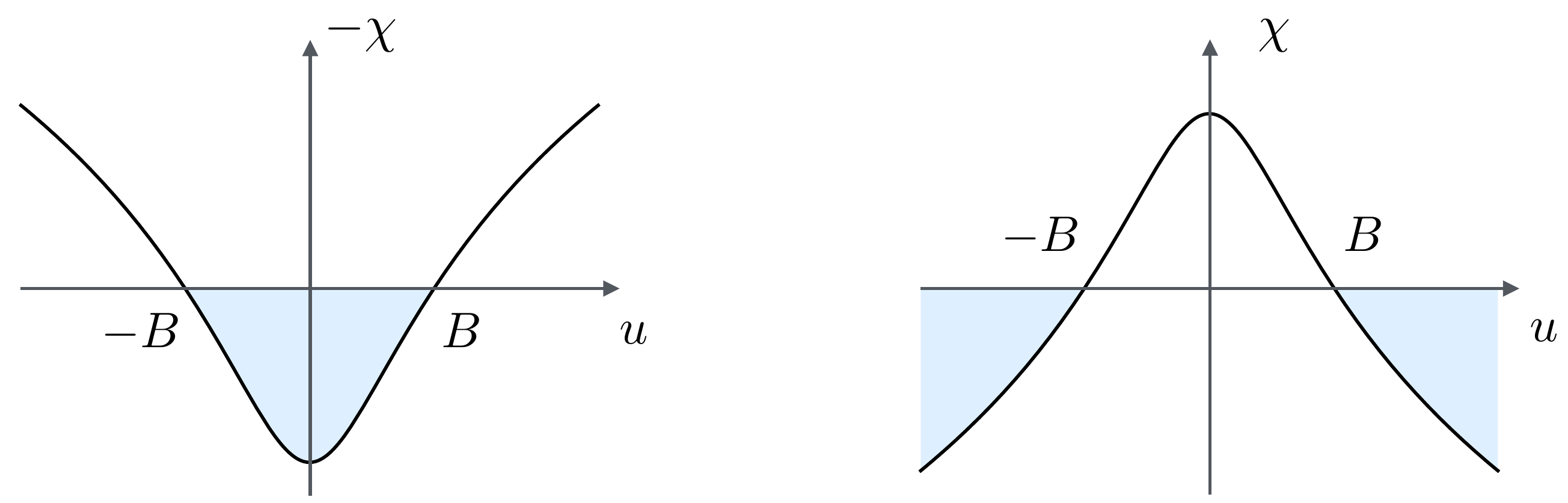}
\end{center}
%\vspace{-7cm}
\caption{Schematic plots of the Fermi sea and its dual. On the left, the original sea, $-\chi \sim h-\epsilon$. Low energies mean small rapidities and all the energy levels inbetween $\pm B$ are occupied. The sea level rises upon increase of the chemical potential $h = \log{g^2}$. On the right, the dual Fermi sea, $\chi\sim E$. Low energies mean large rapidities, the dual excitations accumulate at infinity and all the energy levels beyond the Fermi rapidity are filled. There is no chemical potential here. Instead, the filling is controlled by the charge density $\rho = \Delta/L$, which pilots the large $u$ asymptotics $\chi \sim -\rho \log{u^2}$. When $B\rightarrow \infty$, this density vanishes, the dual sea is empty and the fishnet freezes. This happens at the critical coupling $h = \log{g_{cr}^2}$.}\label{fermi}
\end{figure}

The thermodynamical limit $L\rightarrow \infty$ is uninteresting at weak coupling, since the wheels are heavily suppressed, see equations (\ref{bfYa}) and (\ref{logY}), and the scaling dimension $\Delta = L$ up to exponentially small effects. The situation changes drastically as soon the chemical potential $h = \log{g^2}$ gets bigger than the ``mass gap'' $\epsilon_{1}(0) = \log{1/4}$. Above this threshold the s-wave magnons, with $a=1$, start filling a Fermi sea, see figure \ref{fermi}. The filling is strict in the limit $L\rightarrow \infty$ with all the modes outside the sea being unoccupied,
\beq\label{Y-to-chi}
\log{(1+Y_{1})} \rightarrow L \chi_{1}\theta (B^2-u^2)\, ,
\eeq
with $\theta$ the step function and with the pseudo energy
\beq
\chi_{1}(u) = \lim_{L\rightarrow \infty}\frac{1}{L}\log{Y_{1}}(u) = h - \epsilon_{1}(u) + \ldots ,
\eeq
defined here in such a way that $\chi_{1}\uo 0$ inside/outside the sea. The Fermi rapidity $B$ fixes the edges of the sea and is determined by the condition $\chi_{1}(\pm B) = 0$. Moreover, as long as $g^2$ is smaller than the masses of the higher $a$ magnons, that is, naively, $g^2 < 1$, the gas is mono-atomic and consists solely of fundamental magnons with $a=1$. As we shall see later on we will never reach the next threshold, so in the following we drop the Lorentz index and assume that $a=1$.

An immediate consequence of having a Fermi sea, obtained by plugging (\ref{Y-to-chi}) into (\ref{delta}), is that the scaling dimension develops a non-trivial thermodynamical scaling
\beq\label{f-h}
\Delta/L  = f(h) = 1 - \int\limits_{-B}^{B} \frac{du}{\pi}\, \chi(u)\, ,
\eeq
where $f(h)$ is the free energy density of the gas at chemical potential $h$. The same linearization is observed at the level of the TBA equations (\ref{TBA-1}), see also Section \ref{Sect4}, which can be written concisely as
\beq\label{chi-eq}
\chi(u) = C -\epsilon(u) + \int\limits_{-B}^{B}\frac{dv}{2\pi}\, \mathcal{K}(u-v) \chi(v)\, ,
\eeq
with $\epsilon(u) = \log{(u^2+1/4)}$. Here, for convenience, we split the scattering kernel,
\beq
\frac{\partial}{i\partial u}\log{S(u, v)} = -k(u) + \mathcal{K}(v-u)\, ,
\eeq
into its ``boost'' invariant component, $\mathcal{K}$, and the rest, $k$, which depends on a single rapidity. Their explicit forms follow from equation (\ref{Sab}), with $a=b=1$,
\beq\label{Ku}
\begin{aligned}
\mathcal{K}(u) &=  2\psi(1+iu) + 2\psi(1-iu) + \frac{2}{u^2+1}\, ,\\
k(u) &= 2\psi(\tfrac{1}{2}+iu) + 2\psi(\tfrac{1}{2}-iu)+\frac{1}{u^2+1/4}\, ,
\end{aligned}
\eeq
and with $\psi(z) = \partial_{z}\log{\Gamma(z)}$ the digamma function. Note that both are even functions. The $k$ part of the kernel merely renormalizes the chemical potential and was absorbed into the constant $C$,
\beq\label{Ch}
C = h -\int\limits_{-B}^{B}\frac{du}{2\pi}\, k(u)\chi(u)\, .
\eeq
This system of equations, with the boundary condition $\chi(\pm B) = 0$, admits a unique (even) solution, which can be constructed iteratively for finite value of $B$. The numerical solution for the free energy density is shown in figure \ref{numerics}.

Finally, notice that the free energy density (\ref{f-h}) can also be read out from the large $u$ behaviour of the pseudo energy, which, according to (\ref{chi-eq}) and (\ref{Ku}), scales like
\beq\label{large-u}
\chi(u) =  -f\log{u^2} + C +O(1/u^2)\, .
\eeq
This way of obtaining $f$ will turn out to be useful later on. 

\subsection{Critical coupling}\label{Sect2.3}

Let us now look for Zamoldchdikov's scaling (\ref{Z-th}). First we note that at non zero $B$ the typical number of magnons $T$ in the ensemble is always large, since the gas has a finite particle density,
\beq
j = T/R = -\frac{df}{dh} \neq 0\, .
\eeq
This is necessary for matching with the scaling (\ref{Z-th}), but the condition is not enough to get an actual match. The continuum limit also requires that a low-energy approximation be taken w.r.t.~the fishnet Hamiltonian $\delta/\delta T$. Since, heuristically, $\delta T \sim j$, we expect that the fishnet dynamics will freeze at large magnon densities, that is when $j \rightarrow \infty$. 
Another way of seeing it is that the scaling (\ref{Z-th}) is a statement about the micro-canonical energy density
\beq\label{m-c-b}
\varepsilon = -\frac{1}{RL}\log{Z_{L, T}} \rightarrow j\log{g^2_{cr}}\, ,
\eeq
which, in our thermodynamical variables, translates into the requirement that
\beq
f = \varepsilon - hj \rightarrow 0\, ,
\eeq
for $h \rightarrow \log{g^{2}_{cr}}$. This behaviour cannot be observed at small $B$, that is for a dilute gas, $j\sim 0$, since then $\varepsilon \sim f \sim 1$. On the contrary, the critical regime appears for $B=\infty$, when all the energy levels are filled, which again means that the density is infinite, $j=\infty$.

We can verify it explicitly using the integral equation. Denoting by $\chi_{cr}, C_{cr},$ etc, the limiting values at $B=\infty$, we get to solve
\beq\label{infty-eq}
\chi_{cr} = C_{cr}-\epsilon +\mathcal{K}*\chi_{cr}\, ,
\eeq
where $*$ denotes the convolution over the full real axis,
\beq\label{convolution}
f_{1}*f_{2} = f_{2}*f_{1} = \int\limits_{-\infty}^{\infty} \frac{dv}{2\pi} f_{1}(u-v)f_{2}(v)\, .
\eeq
The solution to (\ref{infty-eq}) can be found by going to Fourier space. Acting on both sides of (\ref{infty-eq}) with
\beq
\int\limits_{-\infty}^{\infty} \frac{du}{2\pi}e^{iut}\partial_{u}\, ,
\eeq
and using the Fourier integrals
\beq
\int\limits_{-\infty}^{\infty}\frac{du}{2\pi}e^{iut}\partial_{u}\epsilon(u)  = \frac{i|t|}{t}e^{-\frac{1}{2}|t|}\, , \qquad \int\limits_{-\infty}^{\infty}\frac{du}{2\pi}e^{iut}\partial_{u}\mathcal{K}(u) = -it\frac{1+e^{-|t|}}{1-e^{|t|}}\, ,
\eeq
immediately tell us that
\beq
\Upsilon(t) = \int\limits_{-\infty}^{\infty}\frac{du}{2\pi} e^{iut}\chi_{cr}(u) = \frac{\sinh{(\frac{1}{2}t)}}{t\cosh{t}}\, .
\eeq
Hence, the free energy density (\ref{f-h}) vanishes
\beq
f_{cr}  =1-\int\limits_{-\infty}^{\infty}\frac{du}{\pi}\chi_{cr}(u) = 1-2\Upsilon(0) = 0\, , 
\eeq
in conformity with our previous discussion. The transformation back to rapidity space yields the critical pseudo energy
\beq\label{chi-cr}
\chi_{cr}(u) = \log{\bigg[\frac{\sqrt{2}\cosh{(\tfrac{1}{2}\pi u)}+1}{\sqrt{2}\cosh{(\tfrac{1}{2}\pi u)}-1}\bigg]}\, ,
\eeq
which is positive definite and decays exponentially quickly at infinity. Plugging it back into (\ref{infty-eq}) and taking a large rapidity limit fix the constant of integration, $C_{cr} = \lim_{u\rightarrow \infty}(\chi_{cr}+\epsilon -\mathcal{K}*\chi_{cr}) = 0$. The critical coupling follows from that condition, after recalling (\ref{Ch}),
\beq
\log{g^{2}_{cr}} = \int\limits_{-\infty}^{\infty}\frac{du}{2\pi} k(u)\chi_{cr}(u) = 2\int\limits_{0}^{\infty}\frac{dt}{t}(e^{-t}-\frac{e^{t}+1}{e^{2t}+1}) =  \log{\bigg[\frac{\Gamma(\tfrac{3}{4})^2}{\pi\Gamma(\tfrac{5}{4})^2}\bigg]}\, ,
\eeq
and it agrees with Zamolodchikov's result (\ref{cr}).

\section{Dualization and AdS sigma model}\label{Sect3}

Having reached the critical point, we want now to study its neighbourhood, corresponding to a large but finite Fermi rapidity $B$. Since ``almost all'' of the energy levels are filled, it is convenient to analyze this regime by means of a particle-hole transformation, which flips the notions of filled and empty states. As we shall see, the dual equation, the one for the holes, is of a totally different nature and lends itself the interpretation of a thermodynamical equation for the tachyon of the hyperbolic sigma model.

\subsection{Particle-hole transformation}

Formally, the particle-hole transformation amounts to introducing the dual kernel
\beq\label{formal-dual-K}
K = -\frac{\mathcal{K}}{1-\mathcal{K}*} = -\mathcal{K}-\mathcal{K}*\mathcal{K}-\ldots\, ,
\eeq
and acting on both sides of (\ref{chi-eq}) with $(1- K*)$. Straightforward algebra gives then
\beq\label{dual-chi-eq}
\chi(u) = I_{dual} + \int\limits_{v^2\geqslant B^2} \frac{dv}{2\pi}K(u-v)\chi(v)\, ,
\eeq
where the convolution is now supported on the complementary support and where $I_{dual} = (1-K*)I$ is the dual of the driving term $I(u) = C-\epsilon(u)$. The procedure is a bit formal since $\mathcal{K}$ scales logarithmically at large rapidity and thus the self-convolutions in the RHS of (\ref{formal-dual-K}) are not well defined. Fortunately, one reaches the same point by defining the dual kernel more implicitly, as the solution to
\beq\label{dual-K}
K =  -\mathcal{K} + \mathcal{K}*K\, .
\eeq
Taking derivative of this equation, going to Fourier space and fixing the constant of integration yield
\beq\label{KO6}
K(u) = \int\limits_{-\infty}^{\infty}dt \frac{e^{|t|}+1}{e^{2|t|}+1} e^{iut} = -i\frac{\partial}{\partial u}\log{S_{O(6)}}(u)\, ,
\eeq
where
\beq
S_{O(6)}(u) = -\frac{\Gamma(1+\tfrac{iu}{4})\Gamma(\tfrac{1}{2}-\tfrac{iu}{4})\Gamma(\tfrac{3}{4}+\tfrac{iu}{4})\Gamma(\tfrac{1}{4}-\tfrac{iu}{4})}{\Gamma(1-\tfrac{iu}{4})\Gamma(\tfrac{1}{2}+\tfrac{iu}{4})\Gamma(\tfrac{3}{4}-\tfrac{iu}{4})\Gamma(\tfrac{1}{4}+\tfrac{iu}{4})} 
\eeq
coincides with the S-matrix of the $O(6)$ sigma model in the symmetric channel \cite{Zamolodchikov:1977nu,Zamolodchikov:1978xm}. It is written here in terms of the Bethe rapidity $u$, which relates to the sigma model hyperbolic rapidity by $\theta = \pi u/2$. Formula (\ref{KO6}) hints at a connection between the dual model and the $O(6)$ sigma model. However, this cannot be the full story, as we shall discuss shortly.

The next essential piece of information comes from $I_{dual}$. Using the normalization of the dual kernel,
\beq\label{norm-K}
\int\limits_{-\infty}^{\infty}\frac{du}{2\pi} K(u) = 1\, ,
\eeq
one concludes that the constant $C$, and hence the chemical potential, drops out of the dual equation, $(1-K*)C = 0$, leaving us with
\beq
I_{dual} = -(1-K*)\epsilon = \chi_{cr}\, ,
\eeq
where we used (\ref{infty-eq}) and (\ref{dual-K}). The dual driving term is thus simply given by the critical pseudo energy (\ref{chi-cr}). It acquires here the meaning of a dual energy, $E = \chi_{cr}$. As noticed earlier, this one decays exponentially at large rapidity. Therefore, the dual energy describes a gapless particle, since one can lower arbitrarily the energy of a dual excitation, by sending it to larger and larger rapidities,
\beq\label{Ed}
E(u) = \log{\bigg[\frac{\sqrt{2}\cosh{(\tfrac{1}{2}\pi u)}+1}{\sqrt{2}\cosh{(\tfrac{1}{2}\pi u)}-1}\bigg]} \sim \frac{m}{2} e^{-\frac{\pi}{2}|u|}\, ,
\eeq
where $m = 4\sqrt{2}$ sets a reference energy scale. This is in line with the fact that the dual Fermi sea has a non-compact support, see figure \ref{fermi}. The dual low energy modes accumulate at infinity, which is typical for gapless systems, see e.g.~\cite{Fateev:1992tk,Zamolodchikov:1992zr,Fendley:1993wq,Fendley:1993xa,Fendley:2000bw,Mann:2004jr}.

It is also natural in the dual picture to exchange the roles of the thermodynamical quantities. The lack of a dual chemical potential, for instance, invites us to view the free energy density $f$ as being part of the specification of the state. Namely, we can simply think of $\rho = f  = \Delta/L$ as the charge density that pilots the large $u$ behaviour,
\beq
\chi \sim -\rho\log{u^2}\, .
\eeq
One can also say that it triggers the formation of the dual Fermi sea, as $\rho$ and $B$ play interchangeable roles.%
\footnote{This is easily seen at the level of the equation (\ref{prime}) for the derivative of $\chi$. Its solution is uniquely fixed at any given $B$ and so is the relation $\rho = \rho(B)$, with $\rho$ being defined by $\chi' \sim -2\rho/u$ at large $u$.  The constant $C$, which appears in the subleading large $u$ behaviour of $\chi$, see eq.~(\ref{large-u}), is then determined by integrating $\chi'$ and imposing that $\chi(\pm B) = 0$.}
In particular, the approach to the critical point $B\rightarrow \infty$ corresponds to the low density regime $\rho\rightarrow 0$.

The last important quantity is the 4d coupling $\log{g^2}$, which, at the moment, is buried inside the constant $C$. Fortunately, one can substitute to (\ref{Ch}) the more transparent relation
\beq\label{marginality}
\log{g^2} = \log{g^{2}_{cr}} +\int\limits_{u^2\geqslant B^2} \frac{du}{2\pi}\, \p_{cr}(u)\chi(u)\, ,
\eeq
where
\beq
\phi_{cr} = -(1-K*)k = \frac{\sqrt{2}\pi \cosh{(\tfrac{1}{2}\pi u)}}{\cosh{(\pi u)}}
\eeq
is the critical micro-canonical distribution density, which is Legendre conjugated to $\chi_{cr}$, and solution to the integral equation
\beq\label{p-cr}
\p_{cr} = -k + \mathcal{K}*\p_{cr} \, .
\eeq
Formula (\ref{marginality}) is obtained by integrating both sides of the original equation (\ref{chi-eq}) against $\phi_{cr}$,
\beq
\begin{aligned}
\int\limits_{-\infty}^{\infty}\frac{du}{2\pi} \phi_{cr}(u)\chi(u) &= C - \log{g_{cr}^2} +\int\limits_{-B}^{B}\frac{du}{2\pi} \chi(u) \mathcal{K}*\phi_{cr}(u) \\
&= C - \log{g_{cr}^2} +\int\limits_{-B}^{B}\frac{du}{2\pi} \chi(u) \phi_{cr}(u) + \int\limits_{-B}^{B}\frac{du}{2\pi} \chi(u) k(u) \\
&= \log{g^2} - \log{g_{cr}^2} +\int\limits_{-B}^{B}\frac{du}{2\pi} \chi(u) \phi_{cr}(u) \, .
\end{aligned}
\eeq
Here, in the first line, we used
\beq\label{id-phi-cr}
\int\limits_{-\infty}^{\infty}\frac{du}{2\pi}  \p_{cr}(u) = 1\, , \qquad \int\limits_{-\infty}^{\infty}\frac{du}{2\pi}  \p_{cr}(u) \epsilon(u) = \log{g_{cr}^2}\, ,
\eeq
and transferred the action of $\mathcal{K}$ from $\chi$ to $\phi_{cr}$, then applied (\ref{p-cr}) in the second line and finally used (\ref{Ch}).

Introducing then a dual momentum $P$, by means of a $90^{\textrm{o}}$ rotation $\theta \rightarrow \theta+i \pi/2$ of the energy,
\beq\label{dual-P}
P(u) = -iE(u+i) = i\log{\bigg[\frac{\sqrt{2}\sinh{\theta}+i}{\sqrt{2}\sinh{\theta}-i}\bigg]} \sim \mp \frac{m}{2}e^{-|\theta|}\, ,
\eeq
and noting that $\phi_{cr} = P' = \partial_{u}P$, one concludes that in the dual description the coupling of the 4d theory is simply related to the 2d energy of the state through the formula (\ref{log-g}) with, in the thermodynamical limit,
\beq
E_{2d}/L = \int\limits_{u\geqslant B^2}\frac{du}{2\pi}P'(u)\chi(u)\, .
\eeq
As we shall see later on, formula (\ref{log-g}) is quite general, and not restricted to the thermodynamical limit.

\begin{figure}
\begin{center}
\includegraphics[scale=0.6]{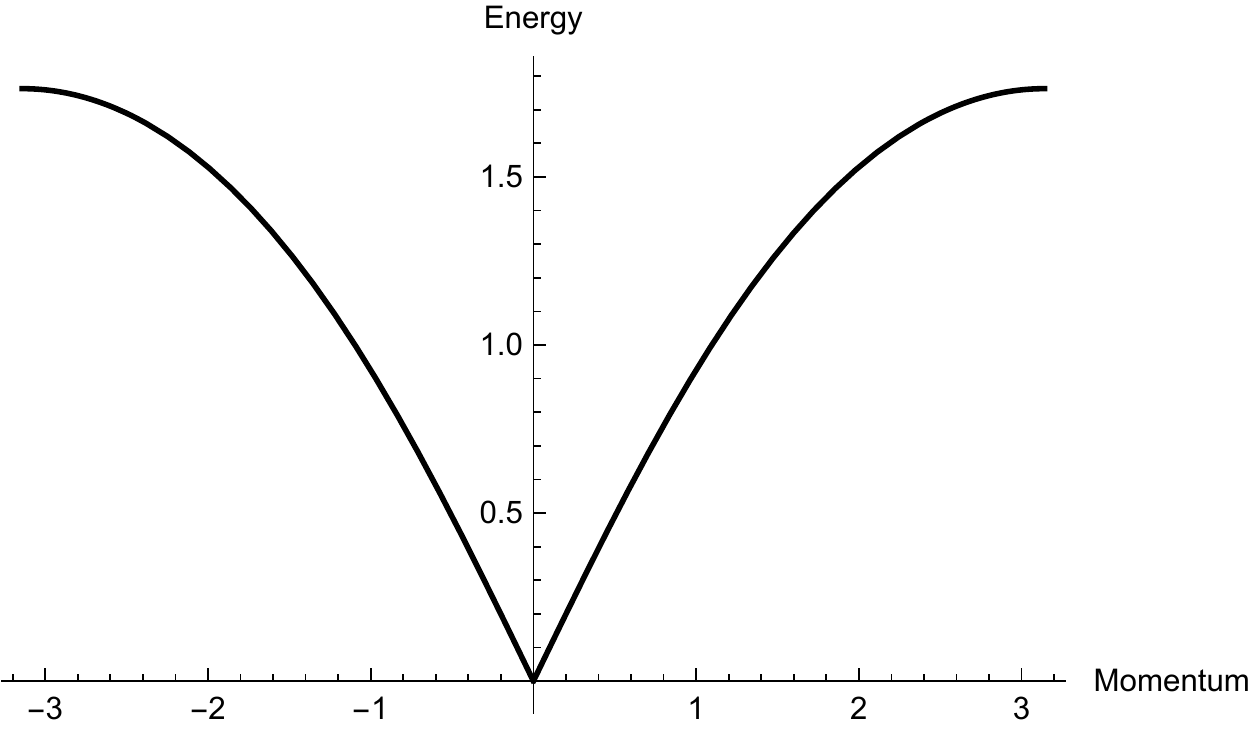}
\end{center}
%\vspace{-7cm}
\caption{Plot of the dual energy $E(P)$ for $P\in (-\pi, \pi)$. At low momentum, we have a massless spectrum $E\sim |P|$. The energy reaches a maximum at $P=\pm \pi$.}\label{dispersion}
\end{figure}

Finally, let us add that putting energy and momentum together yields the dual dispersion relation, which reads
\beq\label{disp}
\sinh^{2}{(\tfrac{1}{2}E)} = \sin^2{(\tfrac{1}{2}P)} \, ,
\eeq
after eliminating the parametric dependence on the rapidity in (\ref{dual-P}) and (\ref{Ed}). This formula makes the square lattice and its symmetries manifest: it preserves a $\mathbb{Z}_{4}$ subgroup of euclidean rotations, generated by $E\leftrightarrow iP$, and shows a maximum momentum, $P = \pm \pi$, and a maximum energy, $E = 2\log{(1+\sqrt{2})}$. Both features disappear at low momentum where one recovers the dispersion relation for a massless relativistic particle. The dispersion relation is depicted in figure \ref{dispersion}, in the 1st Brillouin zone. It is analogous to the energy of a spinon above the anti-ferromagnetic vacuum of the XXX spin chain, and, following this analogy, we could say that at the critical point the spin chain settles down in its symmetric vacuum.

\subsection{Sigma model interpretation}

Let us come to the interpretation of the dual equations. They are very similar to the thermodynamical equation for a (zero temperature) finite density gas of particles, carrying maximal $U(1)$ charge, in the non-linear $O(6)$ sigma model \cite{Hasenfratz:1990ab},
\beq\label{O6-eqs}
\chi(u) = m\cosh{\theta} + \int\limits_{-B}^{B}\frac{dv}{2\pi} K(u-v)\chi(v)\, , \qquad E/L = \int\limits_{-B}^{B}\frac{du}{4}m\cosh{\theta} \chi(u)\, ,
\eeq
where $\theta = \pi u/2$. The difference only comes from the dispersion relation which, in the relativistic low momentum approximation, amounts to substituting
\beq\label{energy-change}
m\cosh{\theta} \rightarrow \frac{m}{2} e^{-|\theta|}\, ,
\eeq
and reversing the support of the distribution. It makes a big difference for the interpretation. In the $O(6)$ model, the particles are massive and though the scattering kernel is repulsive the particles remain confined on a compact support. The kernel has the same effect in our case but the energy is not bounded from below and the potential runs away. Therefore, at finite charge density $\rho$, the excitations start filling the energy levels around $u= \pm \infty$ and spread in the opposite directions, towards smaller rapidities. Also, in the compact case, the charge density $\rho$ matches with the particle density, obtained by integrating the distribution $\chi$ over its support. In our case, the support is non-compact and the distribution is not normalizable, suggesting that the gapless excitations in the condensate cannot be counted.

Given the symmetries of our problem, the most natural guess is that we are dealing with the $AdS_{5}$ sigma model. This non-compact model is known for not developing a mass gap and for having a continuous spectrum in finite volume. Furthermore, the change (\ref{energy-change}) in the energy for a given scattering kernel $K$ embodies the ``inversion of the RG flow'', which is the formal perturbative way of relating the sphere and the hyperbolic sigma model. We discuss it in more detail below.

Finally, note that there are similarities with the equations obtained for massless factorized scattering theories \cite{Fateev:1992tk,Zamolodchikov:1992zr,Fendley:1993wq,Fendley:1993xa}, see also \cite{Fendley:2000bw,Mann:2004jr}. In our case, since we cannot enumerate the particles in the condensate, it is not clear whether we can talk about an S matrix. Put differently, we do not think of our equation as describing the continuum limit of a dense but fundamentally discrete distribution of Bethe roots. On the contrary, the distribution is fundamentally continuous, and will remain continuous after introducing finite size corrections. It defines a one-parameter family of ground states, labelled by $\rho$, or better $\Delta$.

\subsection{Perturbative analysis}

We can verify the interpetation of the dual equation by comparing its predictions against a direct finite density calculation in the sigma model. This is standard analysis for sigma models. It was carried out through two loops in \cite{Hasenfratz:1990ab,Bajnok:2008it} for the sphere. We can follow the same lines for the hyperboloid. In fact, the results for the sphere carry over to the hyperboloid, since compact and non-compact models only differ perturbatively in the sign of the coupling constant, as expected on geometrical grounds \cite{Polyakov:2001af,Friess:2005be,Duncan:2007vs}. We recall how this comes about below.

We consider the non-linear sigma model in Minkowskian $AdS_{d+1}$, where $d=4$ in our case. Its 2d action is given by
\beq
\mathcal{S} = -\frac{1}{2e^2}\int\limits_{-\infty}^{\infty}d\tau\int\limits_{0}^{L}d\sigma\,  \partial^{\alpha} X^{A}\partial_{\alpha} X_{A}\, ,
\eeq
where the embedding coordinates $X^{A} = X^{A}(\sigma, \tau)$ take values on the hyperboloid
\beq
X^{A} X_{A} = -X_{0}^2 + \vec{X}^{2} - X_{d+1}^2 = -1\, ,
\eeq
and where $\vec{X} = (X_{1}, \ldots , X_{d})$ are $d$ transverse directions. The worldsheet metric is taken to be flat, with signature $(-+)$, and we assume periodic boundary conditions in $\sigma$, $\sigma \cong \sigma+L$. The coupling $e^{2}$ sets the curvature of the hyperboloid. In stringy notation $e^{2} = 2\pi \alpha'$ and the model is weakly coupled when $e^2 \sim 0$. The theory has exact $SO(2, d)$ symmetry, with associated conserved Noether currents
\beq\label{Noether}
(J_{\alpha})^{\, A}_{\,\,\, B} = (X^{A}\partial_{\alpha}X_{B}-X_{B}\partial_{\alpha}X^{A})/e^2\, ,
\eeq
and $A, B = 0, \ldots, d+1$.

The model is classically integrable and presumably quantum integrable, perturbatively, for the arguments supporting the integrability of the sphere \cite{Goldschmidt:1980wq} also work for the hyperboloid. The integrability of the non-compact model remains puzzling at the quantum level however. The model's spectrum has no good particle interpretation and thus cannot be handled by the conventional factorized scattering methods. On top of that the model has well known problems in the UV and might have to be completed. However, none of these complications really is a limitation here, as we do not need particles to make use of our equation and we have a lattice to make sense of the model at short distances. Nonetheless, it is interesting to note that both difficulties somehow relate to the running of the coupling $e^2$, which, owing to the negative curvature of the AdS space, is governed by a positive beta function, see~\cite{Friess:2005be} for a recent discussion,
\beq\label{beta}
\mu\frac{\partial e^{2}}{\partial \mu} = \frac{de^{4}}{2\pi}  -\frac{de^{6}}{4\pi^2} +O(e^8)\, .
\eeq
It says that perturbation theory can be trusted at low momentum and this is more than enough for what we intend to do here.

\begin{figure}
\begin{center}
\includegraphics[scale=0.6]{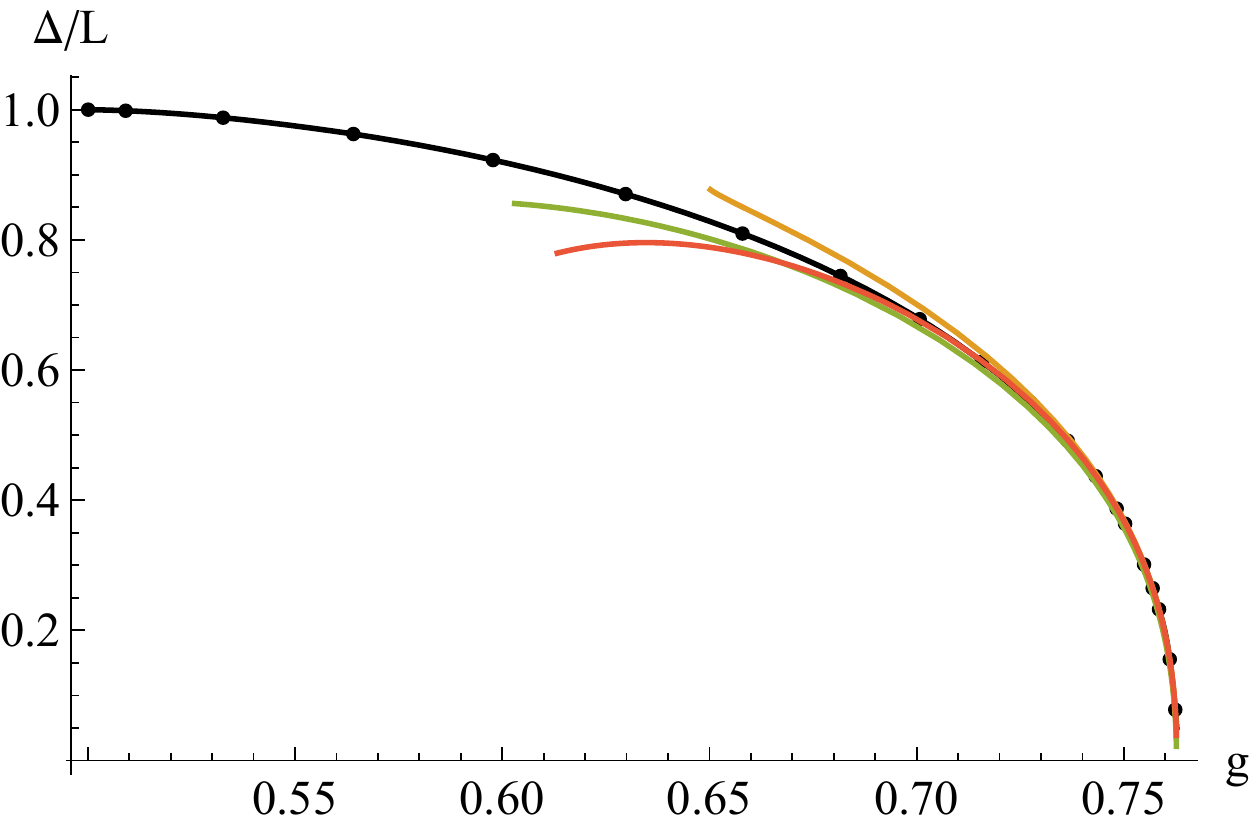}
\end{center}
%\vspace{-6cm}
\caption{Plot of the scaling dimension $\Delta$ as function of the coupling in the thermodynamical window $0.5 <g< g_{cr} \simeq 0.76$. Black dots are numerical values obtained by solving the integral equation using \textsc{Mathematica}. The black line is a numerical interpolation. The green, orange and red lines are perturbative expressions truncated at 1st, 2nd and 3rd order, respectively. They correspond to the classical, one- and two-loop approximation in the sigma model, with the relation $\log{g^2/g^2_{cr}} = E_{2d}/L$.}\label{numerics}
\end{figure}

Getting back to our problem, we seek a state with a charge $\Delta$, along the global time direction, which is uniformly distributed along $\sigma$,
\beq
\Delta/L = (J_{\tau})^{\, 0}_{\,\,\, d+1} \, ,
\eeq
with $J_{\alpha}$ the Noether current (\ref{Noether}), and which corresponds to a local extremum of the sigma model energy,
\beq
E_{2d}/L = \frac{e^2}{2}(J_{\tau})^{\, A}_{\,\,\,B} (J_{\tau})^{\, B}_{\,\,\,A} \, .
\eeq
The most natural candidate is the ``tachyon'', that is, a point-like and time-like geodesics at the center of AdS. It is given classically by the $\sigma$ independent solution
\beq\label{tachyon}
X_{cl}^{0}\pm iX_{cl}^{d+1} = e^{\pm iH\tau}\, , \qquad \vec{X}_{cl} = 0\, ,
\eeq
or, equivalently, $t_{cl} = H\tau$, where $t$ is the global time coordinate. Owing to the signature of the AdS space, ``excitations'' along the time-like direction contribute negatively to the sigma model energy. This applies to our state that has classical energy and charge density
\beq
E_{2d}/L = -\frac{H^2}{2e^{2}} \, , \qquad \rho = \Delta/L = -\frac{H}{e^2}\, ,
\eeq
with a frequency $H$ which is negative for $\Delta$ positive. Eliminating $H$, we obtain that the energy is quadratic in $\rho$,
\beq
E_{2d}/L = -e^2\rho^2/2\, ,
\eeq
up to the running of the coupling. Both the sign and the shape are in agreement with what we observe numerically from the solution to the linear integral equation shown in figure \ref{numerics}. As we shall see, the agreement gets even better when the running of the coupling and the perturbative corrections are included.

Loop corrections are more efficiently computed by exploiting the thermodynamical nature of the state under consideration. Like for the sphere \cite{Hasenfratz:1990ab}, one can access to the energy density by coupling the model to a constant electric field $H$ and extremizing the path integral over an Euclidean worldsheet. Unlike the sphere and owing to the indefinite signature of the target space, one must be careful with the Wick rotation. The kinetic term of the global time coordinate comes out with the wrong sign. We evade the problem by rotating the global time coordinate along with the worldsheet one,%
\footnote{This is automatic, classically, $t_{cl} = H\tau$. We assume that we can also do it for the fluctuations, $\delta t = -i\epsilon$.}
\beq
t\rightarrow -it\, , \qquad \tau\rightarrow -i\tau\, .
\eeq 
It brings us to a doubly euclideanized partition function, which is perturbatively well defined to any order in $e^2$. It remains then to covariantize the $\tau$ derivatives,
\beq\label{CDs}
-\frac{1}{2e^2}(\dot{X}^{0})^2+\frac{1}{2e^2}(\dot{X}_{E}^{d+1})^2 \rightarrow -\frac{1}{2e^2}(\dot{X}^{0}-HX^{d+1}_{E})^2 + \frac{1}{2e^2}(\dot{X}_{E}^{d+1}-HX^{0})^2\, ,
\eeq
with $X_{E}^{d+1} = iX^{d+1}$ the Euclidean coordinate, and expand at weak coupling around the center of the space, a.k.a.~Goldstone vacuum,
\beq\label{Xx}
X^{d+1}_{E} = e y\, , \qquad \vec{X} = e \vec{z}\,, \qquad X^{0} = \sqrt{1 + e^2y+e^2\vec{z}^{\, 2}}\, ,
\eeq
where the fields $y$ and $\vec{z} = (z_{1}, \ldots , z_{d})$ are canonically normalized. It gives
\beq\label{LE}
\mathcal{L}_{E} = \frac{H^2}{2e^2} + \frac{1}{2}\big[(\partial_{a} y)^2+ (\partial_{a} \vec{z}\, )^2+H^2 \vec{z}^{\, 2}\big] + \ldots\, ,
\eeq
where $\mathcal{L}_{E}$ is the Euclidean Lagrangian density and where the dots stand for cubic and higher couplings. The first term in (\ref{LE}) is the classical free energy density $F(H)/L$. The next one shows that the $d$ transverse excitations $\vec{z}$ acquire a mass $|H|$ in the tachyon background,  as in the compact case \cite{Hasenfratz:1990ab}. In fact, everything is as for the sphere model up to $e\rightarrow ie$: this substution flips the signs below the square root in (\ref{Xx}), resulting in the compactification of $X^0$, and turns the derivatives (\ref{CDs}) into
\beq
\frac{1}{2e^2}(\dot{X}^{0}-iHey)^2 + \frac{1}{2e^2}(e\dot{y}+iHX^{0})^2\, ,
\eeq
which is the canonical way of boosting the system along a big circle $\subset S^{d+1}$, see \cite{Hasenfratz:1990ab,Bajnok:2008it}. Therefore, in agreement with the discussions in \cite{Polyakov:2001af,Friess:2005be,Duncan:2007vs}, the compact and non-compact problems are the same, perturbatively, if not for the sign of the coupling, $e^2\leftrightarrow -e^2$.

At this stage, we could import the result directly from the perturbative studies of the sphere sigma model \cite{Hasenfratz:1990ab,Bajnok:2008it} by doing the continuation to negative coupling. In particular the one loop free energy is the same in the two cases and comes directly from the determinants for the quadratic actions in (\ref{LE}). Their evaluation using dimensional regularization gives, in the $\overline{MS}$ scheme,
\beq\label{one-loop}
F(H)/L = \mathcal{h}\mathcal{L}_{E} (H)\mathcal{i} = \frac{H^2}{2e^2} + \frac{dH^2}{4\pi}(\log{(\frac{\mu}{|H|})}+\frac{1}{2}) + O(e^2)\, ,
\eeq
where $e^2 = e^2(\mu)$ solves (\ref{beta}),
\beq\label{running}
\frac{1}{e^{2}(\mu)} = \frac{d}{2\pi} \log{(\frac{\Lambda}{\mu})} - \frac{1}{2\pi}\log{\log{(\frac{\Lambda}{\mu})}} + o(1)\, ,
\eeq
with $\Lambda$ the $\overline{MS}$ scale. Taking the running of the coupling into account, the free energy (\ref{one-loop}) is verified to be independent of the subtraction scale $\mu$. The energy and charge of the state are obtained by a Legendre transformation,
\beq
E_{2d} = F +H \Delta\, , \qquad \Delta = -dF/dH\, ,
\eeq
which, after plugging (\ref{one-loop}), eliminating the coupling (\ref{running}) and fixing $\mu =  2\pi\rho/d$, yields
\beq\label{E-pert}
E_{2d}/(\pi L \rho^2) = -\frac{1}{d\log{(\Lambda/\mu)}} + \frac{(d-1)\log{\log^2{(\Lambda/\mu)}}+d}{2d^2\log^{2}(\Lambda/\mu)} + \ldots.
\eeq
The two-loop calculation \cite{Bajnok:2008it} would also give us the next contribution $\sim 1/\log^3{(\Lambda/\mu)}$. Importantly, since the coupling has disappeared, the same formula (\ref{E-pert}) applies to both the sphere and the hyperboloid. The sole difference is that the expansion is valid for $\rho\ll \Lambda$ in the case the hyperboloid and for $\rho\gg \Lambda$ in the case of the sphere. This is the ``inversion of the RG flow'' alluded to before.

Equation (\ref{E-pert}) can be compared with the formula (\ref{E-int}) obtained from the perturbative analysis of the integral equation done in Appendix \ref{App1}. They agree for $d=4$ if
\beq\label{L-to-m}
\Lambda/m = \left(\frac{e}{8}\right)^{\frac{1}{4}}\Gamma\big(\frac{5}{4}\big)\, ,
\eeq 
which matches with the $\Lambda$-to-$m$ ratio of the $O(6)$ model \cite{Hasenfratz:1990ab}. We recall that in our case $m$ is not a mass gap.

More generally, as shown in Appendix \ref{App1}, our integral equation turns out to be identical at large $B$ with the equation (\ref{O6-eqs}) for the compact sigma model, after continuing $B$ to $-B$. Since the Fermi rapidity plays the role of the radius of the AdS space, changing its sign has the same effect as changing the sign of the coupling $e^2$. Hence, we can bypass the direct comparison with the perturbative sigma model and, assuming the validity of the equation (\ref{O6-eqs}) for the $O(6)$ sigma model, conclude that our equation describes the tachyon of the $AdS_{5}$ model to all orders in perturbation theory.

\section{TBA equations and central charge}\label{Sect4}

The finite density equation only probes a diagonal subsector of the 2d theory and as such might miss some features of the model. More compelling evidence for our proposal can be found by looking at the finite size corrections. We will see that the observations made earlier at the linearized level uplift to the full set of TBA equations. We will then discuss briefly the finite size corrections to the tachyon energy level.

\subsection{Massive TBA}

First recall the original form of the TBA equations for the ferromagnetic vacuum $\textrm{tr}\, \phi_{1}^L$. It is obtained by taking the fishnet limit of the system of TBA equations for the ground state of twisted $\mathcal{N}=4$ SYM spin chain~\cite{Ahn:2011xq}. The relevant symmetry group is the Lorentz group $O(4)$ and the relevant excitations describe the Lorentz harmonics of the scalar field $\phi_{2}$, introduced in Section \ref{Sect2}. These modes appear on an equal footing in the TBA with each mode mapping to a massive (non-relativistic) magnon with bare energy $\epsilon_{a}$, see (\ref{eps-a}), and thermodynamical weight $Y_{a}$, see (\ref{logY}). These Y functions are subject to the equations
\beq\label{gappedTBA1}
\log{Y_{a}} = C-L\epsilon_{a} + \sum_{b\geqslant 1}\mathcal{K}_{a, b}*L_{b} +\sum_{b\geqslant 2} K_{a, b}*(L_{b, +}+L_{b, -}) \, ,
\eeq
where $L_{\star} = \log{(1+Y_{\star})}$ and $*$ denotes the convolution, defined in (\ref{convolution}). The kernel $\mathcal{K}_{a,b}$ controls the part of the interaction that depends on the difference of rapidity, see (\ref{Sab}),
\beq\label{mainK}
\mathcal{K}_{a, b}(u) = \mathcal{K}_{b, a}(u) = \frac{\partial}{i\partial u}\log{\bigg[\frac{\tfrac{a+b}{2}+iu}{\tfrac{a+b}{2}-iu}\frac{\Gamma(\tfrac{|a-b|}{2}+iu)}{\Gamma(\tfrac{|a-b|}{2}-iu)}\frac{\Gamma(1+\tfrac{|a-b|}{2}+iu)}{\Gamma(1+\tfrac{|a-b|}{2}-iu)}\bigg]}\, ,
\eeq
and $\mathcal{K}_{1,1} = \mathcal{K}$, with $\mathcal{K}$ the kernel used before, see (\ref{Ku}). The driving term in (\ref{gappedTBA1}) is given in terms of the bare energy (\ref{eps-a}) and of a constant $C$, which does not depend on the mode number $a$. The latter constant captures the dependence on the coupling constant of the 4d theory and absorbs the part of the kernel that depends on a single rapidity. It reads
\beq\label{full-C}
C = L\log{g^2} -\sum_{a\geqslant 1}\int \frac{du}{2\pi} k_{a}(u) \log{(1+Y_{a})}\, ,
\eeq
where
\beq\label{ka}
k_{a} = 2\psi(\tfrac{a}{2}+i u)+2\psi(\tfrac{a}{2}-i u)+\frac{a}{u^2+\tfrac{a^2}{4}}
\eeq
generalizes (\ref{Ku}) to $a\geqslant 1$.

The last sum in the RHS of (\ref{gappedTBA1}) describes the couplings to the matrix degrees of freedom, represented in the form of $O(4) \sim SU(2)_{+}\times SU(2)_{-}$ dispersion-less magnons, with wave functions $Y_{a, \pm}$. They are labelled by the dimensions $a = 2,3,...$ of the $SU(2)$ representations. The interactions between momentum carrying magnons and isotopic ones are controlled by the scattering kernels of the XXX spin chain,
\beq\label{spin-kernel}
K_{a, b}(u) = K_{b, a}(u) = \sum_{j=(|a-b|+1)/2}^{(a+b-3)/2}\frac{2j}{u^2+j^2}\, .
\eeq
Note in particular that $K_{1, b} = 0$, meaning that there is no coupling between the s-wave magnons and the isotopic ones, as expected.

The TBA equations for the isotopic magnons are entirely controlled by the kernels (\ref{spin-kernel}) and read
\beq\label{o4-wings}
\log{Y_{a, \pm}} = -\sum_{b\geqslant 2}(K_{a, b+1}+K_{a, b-1})*L_{b, \pm} +\sum_{b\geqslant 2} K_{a, b}*L_{b} \, ,
\eeq
with $a\geqslant 2$. For the state we are interested in, there is a left-right symmetry resulting in $Y_{a, +} = Y_{a, -}$.

The scaling dimension $\Delta$ of the BMN operator relates to the free energy of this polyatomic gas and was given in (\ref{delta}). Alternatively, it can be read out from the large $u$ asymptotics of the massive Y functions,
\beq\label{BCs}
\log{Y_{a}} \sim -2\Delta\log{u} + C + \log{m_{a}} + O(1/u^2)\, ,
\eeq
which follows from the universal logarithmic scaling of the kernels $\mathcal{K}_{a, b} \sim 4\log{u}$ and energies $\epsilon_{a} \sim 2\log{u}$. The non-universal part $\log{m_{a}}$ of the sub-leading behaviour comes from the isotopic Y functions $Y_{a, \pm}$. The latter tend at large rapidity to the constant solution of the $O(4)$ Y system, which will be given later on, see equation (\ref{O4}). Plugging this solution into (\ref{gappedTBA1}) and using the normalization of the XXX kernels,
\beq
\int\limits_{-\infty}^{\infty}\frac{du}{2\pi} K_{a, b}(u) = \frac{1}{2}(a+b-|a-b|-2)\, ,
\eeq
one recovers that $m_{a} = a^2$ gives the dimension of the $a$-th Lorentz representation, see (\ref{bfYa}). The relation between global quantum numbers and large rapidity asymptotics is well known in the spin chain context, see e.g.~\cite{BS05}, and plays an essential role in the Quantum Spectral Curve \cite{Gromov:2013pga}.

Equipped with the full set of TBA equations we can verify the claims made earlier about the thermodynamical limit $L\rightarrow \infty$ for $1/2\leqslant g\leqslant g_{cr}$. In particular, we can check that the higher modes $Y_{a>1}$ stay under control in the presence of the condensate $\log{Y_{1}} \sim L\chi $, all the way to the critical point where the back reaction is maximal. Plugging the critical pseudo energy, $\log{Y_{1}} = LE \gg 1$, inside (\ref{gappedTBA1}) and using the identity (\ref{id}) reveal that the functions $Y_{a>1}$ become $L$ independent at the critical point, see (\ref{O6}). This behaviour is indicative of a symmetry enhancement and is further discussed below. Nonetheless, it does not change the fact that the higher modes are negligible thermodynamically. Also, the isotopic Y functions do not couple to the length $L$, nor to $Y_{1}$, and though their actual values depend on $g$ they remain of order $O(L^0)$ all the time, see (\ref{O4}) and (\ref{O6}). Therefore, in the thermodynamical limit one can legitimately substitute $L_{b} \rightarrow \delta_{b, 1} \log{Y_{1}} \theta(B^2-u^2) $ and $L_{b, \pm} \rightarrow 0$ inside (\ref{gappedTBA1}) and this way recover the linear integral equation (\ref{chi-eq}).%
\footnote{Due to an unfortunate choice of notations, in the thermodynamical limit $C_{here}\rightarrow L C_{there}$ where $C_{there}$ refers to the constant (\ref{Ch}).}

\subsection{Massless TBA}

The dualization introduced earlier can be applied to the TBA equations. Let us start with the dual singlet equation, which extends (\ref{dual-chi-eq}) beyond the thermodynamical limit. As before, it follows from acting with $(1-K_{O(6)}*)$ on both sides of the $a=1$ TBA equation (\ref{gappedTBA1}). Then, combining
\beq\label{two}
\mathcal{K}_{1,b}-K_{O(6)}*\mathcal{K}_{1,b} = -K_{2, b}-\delta_{b, 1} K_{O(6)}\, ,
\eeq
which is $b\neq 1$ version of the duality equation (\ref{dual-K}), with the already met relation
\beq
(1-K_{O(6)}*)(C-L\epsilon) = (1-K_{O(6)})*(-L\epsilon) = LE\, ,
\eeq
one immediately obtains
\beq\label{main}
\log{Y_{1}} = LE -K_{O(6)}*L_{1}' - \sum_{b\geqslant 2} K_{2, b}*L_{b}\, ,
\eeq
where $L_{1}' = \log{(1+1/Y_{1})}$ and where $K_{b, 2}$ is the XXX kernel defined in (\ref{spin-kernel}). The rationalization of the couplings between the singlet and the higher modes is the first indication of the restoration of the full symmetry. Notice also that  the new TBA equation (\ref{main}) is consistent with the logarithmic scaling (\ref{BCs}), thanks to the normalization of the $O(6)$ kernel (\ref{norm-K}).

To complete the picture we also verify that the higher modes become dispersion-less in the new vacuum. Indeed, as shown below, the equations for the $a>1$ modes can be brought to the form
\beq\label{dual-hm}
\log{Y_{a}} = -K_{a, 2}*L_{1}' - \sum_{b\geqslant 2}(K_{a, b+1}+K_{a, b-1})*L_{b} +\sum_{b\geqslant 2} K_{a, b}*(L_{b, +}+L_{b, -}) \, ,
\eeq
which shows no coupling anymore to the length $L$ of the system. Together with the equations (\ref{o4-wings}) for the $O(4)$ magnons, which stay untouched in the dual picture, these equations form an $O(6)$ system of magnons sourced by the ``vector'' node $Y_{1}$.

The derivation of the dual equations (\ref{dual-hm}) relies on the relations
\beq\label{one}
\mathcal{K}_{a, b}- \mathcal{K}_{a, 1}*K_{2, b} = -(K_{a, b+1}+K_{a, b-1}) = -\check{K}_{a, b}\, ,
\eeq
which are valid for $a, b >1$. Together with (\ref{two}), they allow us to re-write the first term in the RHS of (\ref{gappedTBA1}) in the form
\beq\label{dualb}
\sum_{b\geqslant 1}\mathcal{K}_{a, b}*L_{b} = -K_{a, 2}*L'_{1}-\sum_{b\geqslant 2}\check{K}_{a, b}*L_{b} + \textrm{rest}_{a}\, ,
\eeq
where we introduced
\beq\label{resta}
\textrm{rest}_{a} =  -K_{a, 2}*\log{Y_{1}}+\mathcal{K}_{a, 1}*K_{O(6)}*L_{1} +\sum_{b\geqslant 2}\mathcal{K}_{a, 1}*K_{2, b}*L_{b}\, .
\eeq
It remains then to show that
\beq\label{want}
\textrm{rest}_{a} =   L\epsilon_{a}-C\, ,
\eeq
which is the statement that the driving term cancels out after shifting to the new vacuum. This can be done in two steps: $(1)$ evaluate $\sum_{b>1}K_{2, b}*L_{b}$ using (\ref{main}), plug the expression in (\ref{resta}) and simplify the result, such as to get
\beq
\textrm{rest}_{a} = \mathcal{K}_{a, 1}*LE -K_{a, 2}*\log{Y_{1}} -  \mathcal{K}_{a, 1}*(\log{Y_{1}}-K_{O(6)}*\log{Y_{1}})\, ,
\eeq
and $(2)$ apply the identities
\beq\label{id}
\mathcal{K}_{a, 1}*LE = L\epsilon_{a}\, ,
\eeq
and
\beq\label{last}
-\mathcal{K}_{a, 1}*(\log{Y_{1}}-K_{O(6)}*\log{Y_{1}}) = K_{a, 2}\log{Y_{1}} -C\, ,
\eeq
which both hold for $a>1$. This is readily seen to imply (\ref{want}). Let us add that though formula (\ref{id}) is straightforwardly derived, the following one (\ref{last}) requires more attention. It is tempting to open up the brackets in the LHS of (\ref{last}) and apply (\ref{two}). This algebra is not correct however and misses the constant $C$. The problem is that $\mathcal{K}_{a,1}*\log{Y_{1}}$ and $\mathcal{K}_{a,1}*K_{O(6)}*\log{Y_{1}}$ are not separately meaningful, because $\log{Y_{1}}$ and $K_{O(6)}*\log{Y_{1}}$ scale logarithmically at large $u$, and so does the kernel $\mathcal{K}_{a, 1}$. However, the large asymptotic behaviours cancel out in their difference as in the LHS of (\ref{last}). To reproduce (\ref{last}) one can first strip out the problematic part, by writing
\beq
\log{Y_{1}} = \log{Y'_{1}} - \Delta\epsilon +C\, ,
\eeq
where $Y'_{1} = Y_{1}e^{\Delta\epsilon-C} \rightarrow 1+O(1/u^2)$ at large rapidity and then apply (\ref{two}). It yields
\beq
\begin{aligned}
-\mathcal{K}_{a, 1}*(1-K_{O(6)}*)\log{Y'_{1}} &= K_{a, 2}*\log{Y'_{1}} = K_{a, 2}*\log{Y_{1}} + \Delta \epsilon_{a} -C\, ,
\end{aligned}
\eeq
after using that $K_{a, 2}*\epsilon = \epsilon_{a}$ and $K_{a, 2}*C = C$ for $a>1$. Finally, adding 
\beq
\begin{aligned}
-\mathcal{K}_{a, 1}*(1-K_{O(6)}*)(- \Delta\epsilon +C) = -\mathcal{K}_{a, 1}*\Delta E = -\Delta \epsilon_{a}
\end{aligned}
\eeq
gives (\ref{last}) and completes the proof.

The TBA energy is defined canonically by
\beq\label{E-TBA}
E_{2d} = -\int\limits_{-\infty}^{\infty}\frac{du}{2\pi} P'(u) \log{(1+1/Y_{1}(u))}\, ,
\eeq
where $P$ is the dual momentum introduced in (\ref{dual-P}). It is related to the coupling constant of the 4d theory by means of (\ref{log-g}) as found earlier in the thermodynamical limit. The proof is similar to the one outlined below equation (\ref{marginality}). Namely, one integrates both sides of the TBA equation (\ref{gappedTBA1}) for $\log{Y_{1}}$ against $\p_{cr}(u) = P'(u)$, transfer the action of the kernels to $\phi_{cr}$, and use the identities
\beq
\mathcal{K}_{a, 1}*\phi_{cr}(u) = \delta_{a, 1} \phi_{cr}(u) + k_{a}(u)\, ,
\eeq
as well as the definition (\ref{full-C}) and the relations (\ref{id-phi-cr}).

Equations (\ref{main}), (\ref{dual-hm}) and (\ref{o4-wings}) form the dual set of TBA equations. They take the same form as for the ground state of the $O(6)$ sigma model if not for the driving term, which is given by the gapless energy (\ref{Ed}). This feature allows us to impose the boundary condition (\ref{BCs}) at large rapidity, or equivalently small momentum, which in the dual picture is part of the specification of the state. Therefore, as seen before, the TBA equations do not describe a single isolated vacuum, as in the compact case, but a one-parameter family of vacua labelled by the scaling dimension $\Delta$. The sigma model energy (\ref{E-TBA}) is then determined as a function of the system size $L$ and quantum number $\Delta$. Formula (\ref{log-g}) is the only place where the coupling constant of the 4d theory enters.

\begin{figure}[t]
  \centering
    \subfloat{\includegraphics[width=0.4\textwidth]{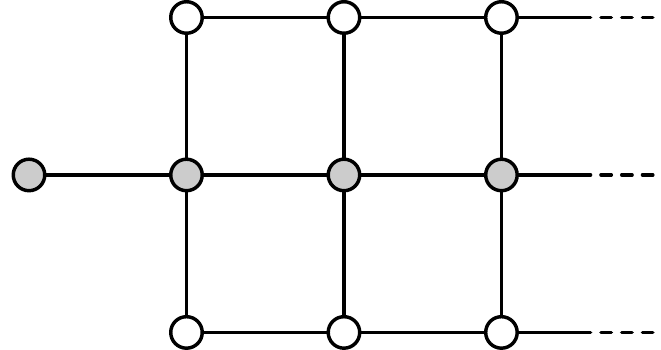}}
    \qquad%
    \subfloat{\includegraphics[width=0.4\textwidth]{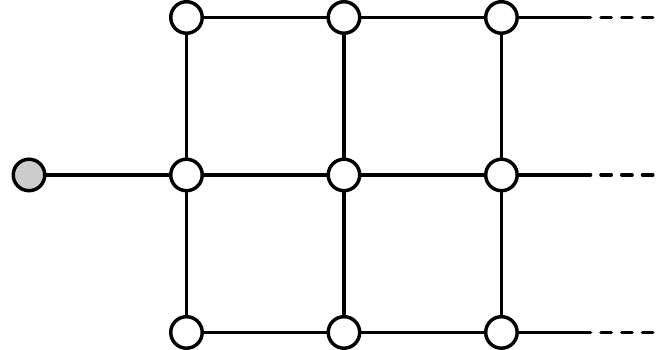}}%
    \vspace{0.7cm}
  \caption{Summary of the massive and dual massless TBA equations, on the left and right panel, respectively. Each node represents a Y function. Filled nodes stand for the momentum carriers while white nodes denote dispersion-less magnons. On the left panel, the momentum carrying nodes form the array $Y_{a}$, with $a=1, 2, \ldots\,$ and with $a=1$ being the singled out node. This array is connected on both sides to an $SU(2)$ system of magnons, if not for $a=1$, which is a Lorentz singlet. In the dual description an $O(6)$ tail of magnons is attached to a single momentum carrying node, which is in the vector of $O(6)$.}\label{TBAs}
\end{figure}

Note finally that although our model is gapless we do not witness the splitting into left and right movers that is characteristic of massless factorized scattering theories \cite{Zamolodchikov:1992zr,Fateev:1992tk}. This is due to the fact that we are dealing with a lattice completion of the sigma model. The 2d boost symmetry is broken at short distances, not at the level of the scattering kernels, which are as in the $O(6)$ model, but because of the dispersion relation, see figure \ref{dispersion}. The latter permits the left and right movers to join up at high energy. Alternative UV completions might differ in the way left and right movers are put together into the TBA equations. It would be interesting to investigate what are the possible options if one insists on having exact relativistic symmetry.

\subsection{Y system}

The two formulations of the TBA equations are summarized in figure \ref{TBAs}. As a final check of their equivalence one can verify that they give rise to the same Y system. The latter refers to functional relations that the Y functions must obey. One verifies, using fusion relations among the TBA kernels, that the $O(6)$ tail of equations takes the universal form
\beq\label{A3}
\frac{Y_{a, s}^{+}Y_{a, s}^{-}}{Y_{a+1, s}Y_{a-1, s}} = \frac{(1+Y_{a, s+1})(1+Y_{a, s-1})}{(1+Y_{a+1, s})(1+Y_{a-1, s})}\, ,
\eeq
for $a>1$ and $s = 0, \pm 1$, while the equation for the singlet, $a=1$, is more model dependent and reads
\beq
\frac{1}{Y_{1}^{++}Y_{1}^{--}} = (1+\frac{1}{Y_{2, +}})(1+\frac{1}{Y_{2}^{+}})(1+\frac{1}{Y_{2}^{-}})(1+\frac{1}{Y_{2, -}})\, .
\eeq
Here, $Y_{a, s = 0} = Y_{a}, Y_{a, s=\pm 1} = Y_{a, \pm}, Y^{\pm}(u) = Y(u\pm \tfrac{i}{2}),$ etc. They are seen to match the Y system equations for the $O(6)$ sigma model~\cite{Balog:2001sr,Balog:2005yz,Fendley:1999gb} up to change of notations.
 
\subsection{IR central charge}

Since our model is gapless, the typical energy scale for the finite size corrections is $1/L$. Hence, the thermodynamical description no longer applies when the charge density $\rho$ becomes comparable to $1/L \ll 1$. On the sigma model side, we can get to this far-infrared domain by sending $L \rightarrow \infty$ at fixed $\Delta$ and then read out the energy level from the 2d scaling dimension of the tachyon vertex operator, using the operator-state correspondence \cite{Polyakov:2001af,Tseytlin:2003ac}. Alternatively, we can push the thermodynamical analysis to the very low density domain $\rho L \sim 1$ by including the finite size corrections to the one loop determinants. We will follow the latter approach here.

Recall that we found 1 longitudinal massless boson and $d$ transverse massive bosons, with a mass $-H$ set by the chemical potential $H$ (which is negative). The sole effect of the massless boson is to contribute one unit of central charge to the free energy
\beq
\delta_{y} F = -\frac{\pi }{6L}\, .
\eeq
The contribution of the $d$ transverse bosons is more interesting and takes the form
\beq\label{perp-F}
\delta_{\vec{z}} F = d\int\limits_{-\infty}^{\infty}\frac{dp}{2\pi} \log{(1-e^{-L\sqrt{p^2+H^2}})} = -\frac{d|H|}{\pi}\sum_{k=1}^{\infty}\frac{1}{k}K_{1}(k|H|L) \, ,
\eeq
where $K_{1}$ is the 1st modified Bessel function of the second kind. It is exponentially small at large $L$, for $|H|\sim 1$, as expected. At very low density, $|H| L \ll 1$, when the bosons become effectively gapless, it can be expanded as
\beq\label{small-H}
\delta_{\vec{z}} F = -\frac{d\pi}{6L} -\frac{dH}{2}+\frac{dH^2L}{4\pi}(\log{(|H|L/4\pi)}+\gamma_{E}-\frac{1}{2}) + O(H^3L^2) \, ,
\eeq
with $\gamma_{E}$ the Euler-Mascheroni constant. Besides the Casimir energy, we find a term linear in $H$, which has the interesting effect of shifting the extremum value of $\Delta$, at $H = 0$,
\beq
\Delta = -\frac{dF}{dH} = \frac{d}{2} + O(H)\, .
\eeq
The next term in (\ref{small-H}) is quadratic in $H$ and resembles the bulk term (\ref{one-loop}). Its role is to switch the argument of the coupling from $|H|$ to $1/L$. Adding up the bulk and finite size contributions and performing the Legendre transformation, one obtains the energy
\beq\label{Casimir}
E_{2d} = -\frac{\pi c}{6L} - \frac{e^{2} \Delta(\Delta-d)}{2L} + \ldots \, ,
\eeq
where $e^{2}\sim 2\pi/(d \log{L})$ and $c = d+1$. This formula formally agrees with the CFT prediction for the tachyon energy \cite{Polyakov:2001af,Friess:2005be,Tseytlin:2003ac}, being the sum of the Casimir energy and of the tachyon kinetic energy, which is yet another Casimir, for the quadratic invariant of the tachyon representation this time. However, as written the formula is not complete and misses the loop correction to the central charge; the latter is of the same order as the kinetic part but originates from the two loop correction to the free energy $F(H)$ at $H=0$. To complete it, one just needs to replace $c$ in (\ref{Casimir}) by the effective central charge~\cite{Friess:2005be}
\beq
c_{\textrm{eff}}(L) = d+1 +\frac{3d(d+1)e^2}{4\pi} + O(e^4)\, .
\eeq
This addition has a minor effect though and to summarize the main points are that $(a)$ the extremum of the sigma model energy is shifted by the finite size effects to $\Delta = d/2$ in agreement with the observations made in \cite{Gromov:2017cja,Grabner:2017pgm} and $(b)$ the coupling $e^2$ never reaches the ``Moscow zero'' and freezes at the IR energy scale $\mu = 1/L\ll 1$.

Notice also that the extremum energy is not simply given by the would-be ``vacuum energy'' $-\pi c_{\textrm{eff}}(L)/6L$ and gets corrected by the kinetic term,
\beq
E_{2d}(\Delta = d/2) = -\frac{\pi(d+1)}{6L}-\frac{e^2 d}{8L} + O(e^4)\, .
\eeq
Inserting this energy into formula (\ref{log-g}) and setting $d=4$ yield the leading finite-size correction to the location of the branch point at large $L$,
\beq
g_{cr}(L) = 0.762(8) -\frac{0.998(5)}{L^2} +O(1/(L^2\log{L}))\, ,
\eeq
which moves closer to the findings of \cite{Gromov:2017cja} and fits remarkably well the finite $L$ numerical points obtained from the Baxter equation.%
\footnote{We are very thankful to David Grabner and Kolya Gromov for sharing with us their findings for the critical coupling at $L=4,5$ and for comparing our asymptotic expression against their numerical results. }

The far-infared domain is significantly harder to study on the TBA side. Although it might be possible to address it using advanced TBA techniques, like the ones developed in e.g.~\cite{Gromov:2009tq,Zamolodchikov:1991pc,Fateev:1992tk,Zamolodchikov:2000kt,Teschner:2007ng}, here we will content ourselves with running the standard dilog routine for computing the central charge \cite{Zamolodchikov:1989cf,Klassen:1989ui,Klassen:1990dx,Zamolodchikov:1991et,Zamolodchikov:1991pc}.

The main simplification that comes about at large $L$ is that the solution splits into decoupled left- and right-moving kinks, centered around $\pm \log{(\frac{1}{2}mL)}$ respectively. Each kink corresponds to a scale invariant solution of the TBA equations and, by parity, the two kinks contribute the same amount to the energy, which takes the scale invariant form $E_{2d} = -\pi c_{TBA}/6L$. Remarkably, the exact shape of the kink appears to be immaterial at first order and the central charge $c_{TBA}$ is directly determined by the boundary values of the Y functions at the extremities of the kink, denoted by $Y^{0}$ and $Y^{\infty}$ in the following. The precise relation is $c_{TBA} = c_0-c_{\infty}$, where
\beq\label{dilog-sum}
c_{\star} = \sum_{i} \mathcal{L}(\frac{Y^{\star}_{i}}{1+Y^{\star}_{i}})
\eeq
is a sum over all the nodes of the Y system and where $\mathcal{L}$ denotes Rogers dilogarithm,
\beq
\mathcal{L}(x) = \frac{6}{\pi^2}(\textrm{Li}_{2}(x) + \frac{1}{2}\log{x}\log{(1-x)})\, .
\eeq
In our case the kink interpolates between the two phases described earlier. Each is characterized by a symmetry group and an associated constant solution to the Y system (\ref{A3}), with vanishing boundary conditions at $a= \infty$. At large $u$ we are deep inside the dual sea, the symmetry is broken down to $O(4)$ and so the relevant solution is the stationary $A_{2}$ solution,
\beq\label{O4}
Y_{a}^{\infty} = 0\, , \qquad Y^{\infty}_{a, \pm} = \frac{1}{a^2-1}\, .
\eeq
On the other hand, fixing the rapidity and sending $L$ to infinity place us far from the dual sea, where the $O(6)$ symmetry is restored. In this case we need the constant $A_{3}$ solution,
\beq\label{O6}
Y^{0}_{1} = \infty\, , \qquad Y^{0}_{a, s} = \frac{4-|s|}{(a-1)(a+3)}\, .
\eeq
(Note, in passing, that this solution is telling us how the Y functions for the higher modes behave close to the critical point; they tend to constants $Y_{a} <1$.) Evaluating the dilog sum (\ref{dilog-sum}) numerically on these two solutions, one infers that
\beq
c_{0} = 7\, , \qquad c_{\infty} = 2\, ,
\eeq
and recovers that $c_{TBA} = 5$, as expected. Since this analysis is not sensitive on the way the Y functions decay at infinity, it does not capture the information about the scaling dimension $\Delta$.

Note finally that the same algebra applies to the compact model, with the same central charge. Generalizing to $d$ dimensions, one expects that a scale invariant solution interpolating between $O(d+2)$ and $O(d)$ boundary conditions will reproduce the central charge of the $O(d+2)$ sigma model. This expectation can be verified numerically for even $d$ using the constant solution to the $D_{r}$ Y system given in \cite{Balog:2001sr} (see also references therein).

This computation relates to earlier TBA analyses for the central charge of the sphere sigma models \cite{Balog:2001sr,Fendley:1999gb}, although the details of the calculations are a bit different. The latter references considered the integrable deformation of the level $k$ coset WZNW model, conjectured in \cite{Fendley:1999gb} to approach the sphere sigma model in the limit $k\rightarrow \infty$. This deformation translates at the level of the Y system into a truncation at $a=k$, where a hard wall $Y_{k, s} = \infty$ is located. Although the kink solutions for the truncated system tend to the undeformed ones when $k\rightarrow \infty$, one finds a remnant of the wall in $c_{0}$ and $c_{\infty}$ even after sending $k\rightarrow\infty$. The latter coefficients indeed receive additional contributions from the wall region $a\sim k$ and the excess only drops out in the physical central charge $c = c_{0}-c_{\infty}$.

It would be interesting to study the truncation of the TBA equations for the non-compact model as well as its field theoretical realization along the lines of \cite{Fendley:1999gb}. A truncated system would certainly be more tractable both numerically and analytically. One puzzling question that comes to mind is whether a truncation can be found for every $\Delta$. Indeed it might not be possible for a truncated system to support the logarithmic behaviour (\ref{BCs}) at $u = \infty$. In fact, looking at the expressions obtained in \cite{Balog:2001sr} for the compact model, one might expect \textit{all} Y functions to approach non-zero constant values at $u=\infty$, for finite $k$, although some of them can be made arbitrarily small at large $k$,
\beq
\log{Y_{a}} \sim -d \log{k}\, .
\eeq
The similarity of this scaling with the $\Delta = d/2$ asymptotic behaviour (\ref{BCs}) suggests that the tachyon will settle down at the extremum of the energy in the $k\rightarrow \infty$ limit. More generally, it could be that the truncation is only possible for certain quantized values of the spectrum and comes along with a certain compactification of the target space.

\section{Conclusion}\label{Sect5}

In this paper, we applied integrability to the study of the thermodynamical limit of the 4d planar fishnet graphs. The general proposal is that the fishnet diagrams correspond to an integrable lattice regularization of the $AdS_{5}$ sigma model. We tested this correspondence perturbatively for the scaling dimension of the BMN operator $\textrm{tr}\,\phi_{1}^L$ which maps to the tachyon on the sigma model side. It would be very interesting to see if the correspondence can be understood within the formalisms of \cite{Gromov:2017cja,Zamolodchikov:1980mb} for these ones relate more directly to the Feynman graphs. 

Our discussion fits with the familiar correspondence between spin chains and sigma models in condensed matter physics, which relates in its emblematic form the antiferromagnetic Heisenberg magnet to the $O(3)$ sigma model at topological angle $\theta = \pi$. Our set up appears orthogonal to the traditional one however, for $(1)$ the symmetry group is non compact, $(2)$ the vacuum is ferromagnetic and $(3)$ the interactions are non local. Ferromagnetic spin chains come usually with a non-relativistic spectrum \cite{Polyakov:2005ss}. What saves us from that fate is that our ``magnet'' loses its ferromagnetic property at the critical point, as indicated by the vanishing of the ``scaling dimension per field'' $\Delta/L$ in the thermodynamical limit. (The latter is an order parameter for the conformal symmetry breaking.) This loss comes along with the emergence of the symmetric phase described by the AdS sigma model. It is not clear to us whether the non locality of the interactions is essential in this respect. In any event it would be interesting to see how generic the correspondence is by considering other integrable fishnets \cite{Zamolodchikov:1980mb,Caetano:2016ydc,Mamroud:2017uyz,Chicherin:2017frs,Kazakov:2018qbr}, with dimension $d\neq 4$, deformed propagators, Yukawa couplings, etc. In particular, the $d=3$ triangular fishnets can be embedded into the ABJM theory \cite{Caetano:2016ydc} and studied along the lines of this paper. The linearized analysis is carried out in Appendix \ref{App3} and hints at a connection with the $AdS_{4}$ sigma model. It would also be interesting to explore the potential connections with the work of \cite{Ikhlef:2011ay} which demonstrates that certain non-compact sigma models can be engineered from the continuum limit of anti-ferromagnetic spin chains with compact symmetry group but non-Hermitian dynamics.

Another important question is whether all the local operators of the theory find room in the sigma model description. Some operators are known to be protected and thus seem to evade the sigma model. The excited operators look more promising. In Appendix \ref{App2} we give evidence that adding spin to the operator is in line with adding spin to the tachyon. A more systematic analysis would be needed for comparison with the full spectral curve of the classical $AdS_{5}$ sigma model, along the lines of \cite{Kazakov:2004qf,Gromov:2006dh}.

The relation (\ref{log-g}) between the 4d coupling $g^2$ and the sigma model energy is also worth a few comments. First one notices that this coupling decouples from the story and only enters in the latter relation to the energy. This is not too surprising given that the fishnet diagrams themselves do not know about it. The latter is inserted in the sigma model in the form of a cosmological constant and connects to the energy because of the summation over the discrete modular time, or Schwinger parameter,
\beq
\sum_{T\geqslant 0} (g/g_{cr})^{2LT} e^{-TE_{2d}(L, \Delta)} = \frac{1}{1-(g/g_{cr})^{2L} e^{-E_{2d}(L, \Delta)}}\, .
\eeq
It explains why the relation to the 4d coupling takes the form of a ``marginality condition'', for the on-shell states of the fishnet theory are associated to the poles of the ``string propagator''. The condition is exact, being valid for any $L$, but only at large $L$ is the sigma model weakly coupled. Nonetheless, even at small $L$ one should be able to write (part of) the spectrum in this implicit form. This is in line with the observation made in \cite{Grabner:2017pgm} for $L=2$, where a four-point function of short operators was cast into the form of an integral over a continuum of scaling dimensions, with physical states sitting at the poles of the integrand. The physicality condition in this case was found to take the simple form
\beq
(g/g_{cr})^{2L} e^{-E_{2d}(L, \Delta)}\big|_{L=2} = \frac{16g^{4}}{\Delta(\Delta-2)^2(\Delta-4)}\, ,
\eeq 
for spin-less operators. More generally, the marginality condition of the sigma model should relate to the eigenvalues of the graph building operator \cite{Gromov:2017cja}. Note that similar representations were also found for the correlators of the SYK model  \cite{Gross:2017aos}, which is graphically a close relative of the fishnet theory.  

It remains to be seen if this on-shell condition admits a genuine stringy interpretation. String worldsheet theories with a free tunable intercept do exist in flat space, if one quantizes the Polyakov action \textit{\`a la} Gupta-Bleuler \cite{Polchinski:1998rq}. These non-critical worldsheets do not seem to lift to consistent interacting string theories, but they seem to be reasonably well behaved classically. An estimate of the asymptotic number of states in the fishnet theory or relatedly of the Hagedorn temperature \cite{Harmark:2017yrv,Harmark:2018red} could shed light on this issue.

It would also be interesting to see what the sigma model can teach us about the structure constants and the higher point functions. It might naturally relate to the integral representations obtained in \cite{Grabner:2017pgm} and \cite{Gross:2017aos}.

Finally, our analysis could also find applications in other critical corners of the $\mathcal{N}=4$ SYM theory. For instance, the critical behaviour of scaling dimensions in the fishnet theory is reminiscent of the tachyonic instabilities discussed in \cite{Bajnok:2013wsa}. Also, the BFKL limit, which relates to the point in the spin plane where the derivative of the scaling dimension diverges, is another possible place where a contact could be made; see \cite{Alfimov:2018cms} for recent discussion. More generally, one might be able to find similitudes with observables that are dominated by generalized ladder dynamics, like the one recently considered in \cite{Cavaglia:2018lxi}.

\section*{Acknowledgments}

We are thankful to Lance Dixon and David Kosower for collaboration on a related dense fishnet project. We also thank Camille Aron, Andrei Belitsky, Joao Caetano, David Grabner, Kolya Gromov, Volodya Kazakov, Shota Komatsu, Grisha Korchemsky, Ivan Kostov, Giuseppe Policastro, Vladimir Rosenhaus, Didina Serban, Amit Sever, Kostya Zarembo and Aleksandr Zheltukhin for useful discussions and Volodya Kazakov, David Kosower and Grisha Korchemsky for many comments on the manuscript. Thanks also to the participants of the workshop on higher-point correlation functions and integrability in AdS/CFT, Dublin April 2018, to the CERN String Theory people and to the participants of the Nordita Program on Correlation Functions in Solvable Models, Stockholm May 2018, for useful feedback. B.B. would like to thank Nordita for hospitality during the last stage of the writing of this paper. This work was supported by the French National Agency for Research grant ANR-17-CE31-0001-01.

\appendix

\section{Low density analysis}\label{App1}

In this appendix we study the dual integral equation at large $B$, corresponding to a low charge density $\rho \ll m$. We shall demonstrate that its solution matches the one for the compact sigma model at every order in the $1/B$ expansion, up to $B\rightarrow -B$. Since $B$ relates to the inverse of the sigma model coupling, switching its sign embodies the change of curvature which maps the sphere into the hyperboloid. A similar rule maps the UV and IR regimes of the finite density equations discussed in \cite{Zamolodchikov:1992zr,Fateev:1992tk} related to marginally relevant and irrelevant (integrable) deformations of $SU(2)_{1}$ WZNW model.

We consider the equation for the derivative of $\chi$, which is technically simpler. It reads
\beq\label{prime}
\frac{d\chi}{du}(u) = \chi'(u) = E'(u) + \int\limits_{\mathcal{C}}\frac{dv}{2\pi} K_{O(6)}(u-v)\chi'(v)\, ,
\eeq
where $\mathcal{C} = \{u\in \mathbb{R} : u^2>B^2\}$ is the support of the dual distribution, $E'$ is the derivative of (\ref{Ed}),
\beq\label{E-prime}
E'(u) = -\frac{\sqrt{2}\pi \sinh{(\tfrac{1}{2}\pi u)}}{\cosh{(\pi u)}} \sim -\frac{\pi m}{4} \textrm{sign}(u) e^{-\frac{\pi}{2}|u|}\, ,
\eeq
with $m=4\sqrt{2}$, and $K_{O(6)}$ is the $O(6)$ kernel (\ref{KO6}). The equation is obtained by differentiating (\ref{dual-chi-eq}), performing an integration by part and using that $\chi(\pm B) = 0$. Notice also that $\chi'$ is odd and scales like
\beq\label{rho-prime}
\chi' \sim -2\rho/u\, ,
\eeq
at large $u$, where $\rho = \Delta/L$ is the charge density. This equation can be analyzed systematically at large $B$ by following closely the method developed by Volin in \cite{Volin:2009wr,Volin:2010cq} for the compact sigma model. The reader is also referred to~\cite{Hasenfratz:1990ab} for an earlier study based on the Wiener-Hopf method. 

To begin with, one introduces the resolvent
\beq\label{Resolvent}
R(u) = \int\limits_{\mathcal{C}} \frac{dv}{2\pi}\frac{\chi'(v)}{u-v}\, ,
\eeq
which is an even analytical function of $u$ if not along the contour $\mathcal{C}$ where it has the discontinuity
\beq\label{R-disc}
iR(u+i0)-iR(u-i0) = \chi'(u)\,  \theta(u^2-B^2)\, ,
\eeq
with $\theta$ the step function. It is required to scale like $R(u) \sim \pm i\rho/u$ when $u \rightarrow \pm i\infty$, such as to fulfill (\ref{rho-prime}).%
\footnote{Because the support is non-compact, the resolvent (\ref{Resolvent}) is not analytical at infinity and must contain an essential singularity $\sim e^{-\frac{\pi}{2} |u|}$ to comply with equation (\ref{R-eq}). However, this non-perturbative singularity is not visible in the $1/B$ expansion for $u/B = O(1)$.} One can solve for the resolvent directly after casting (\ref{prime}) into the form
\beq\label{R-eq}
\frac{1-D}{1+D^2}R(u+i0) - \frac{1-D^{-1}}{1+D^{-2}}R(u-i0) = -iE'\, ,
\eeq
with $u\in \mathcal{C}$ and where $D = e^{i\partial_{u}}$ is the shift operator, $D^{k}f(u) = f(u+ik)$.

\begin{figure}
\begin{center}
\includegraphics[scale=0.45]{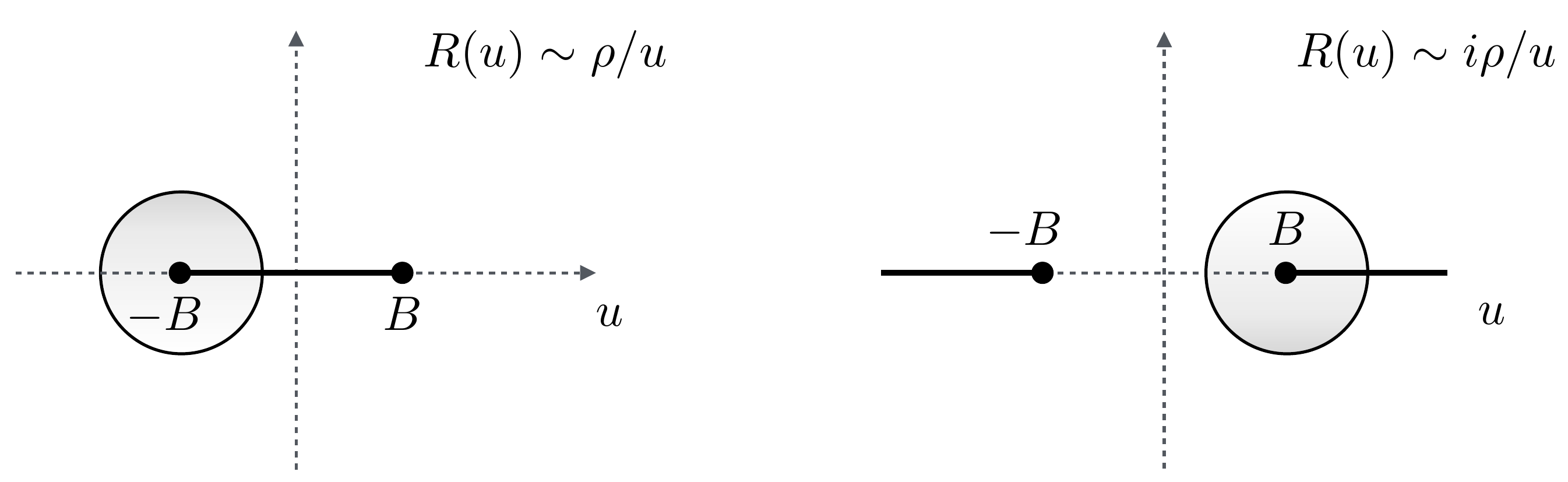}
\end{center}
%\vspace{-8cm}
\caption{Cut planes of the resolvents for the compact and non-compact models, in the left and right panels, respectively. The two problems differ in the positions of their cuts. One can formally relate their perturbative solutions close to the branch point by changing the sign of $B$.}\label{cuts}
\end{figure}

For $u/B = O(1)$ the large $B$ analysis of (\ref{R-eq}) takes the form of a gradient expansion, since then the derivatives are small, $\partial_{u}\sim O(1/B)$, and the shift operator can be expanded, $D = 1+i\partial_{u}+\ldots\,$. The equation reduces then to a Riemann-Hilbert problem, which can be solved iteratively, order by order in $1/B$. At first, one would like to get rid of the driving term, which gives rise to exponentially small corrections in this ``hydrodynamical'' regime. This can be done neatly, following \cite{Volin:2009wr}, by $(1)$ acting on both sides of (\ref{R-eq}) with $(D+D^{-1})$, which explicitly removes the RHS, owing to its $2i$-periodicity, and $(2)$ factorizing out $(D^{1/2}-D^{-1/2})$. Inverting the latter operator perturbatively $\sim -i\partial^{-1}_{u} + O(\partial_u)$ results in a constant of integration which must vanish in our case, by parity. Going along these lines brings the equation
\beq\label{DR}
D^{-1/2} R(u+i0) + D^{1/2} R(u-i0) = 0\, ,
\eeq
which is free of inhomogeneous term, but is nonetheless equivalent to (\ref{R-eq}) to all orders in $1/B$ for $u/B = O(1)$. This is the same equation as for the $O(6)$ sigma model \cite{Volin:2009wr}, since in both cases the driving term drops out after performing the $(D+D^{-1})$ projection. The only difference resides in the domain of definition, which is non-compact in our case, $u\in \mathcal{C}$, see figure \ref{cuts}.

To leading order, $D\rightarrow 1$, and the problem reduces to a well-known singular equation,
\beq
R_{0}(u+i0)+R_{0}(u-i0) = \dashint\limits_{\mathcal{C}}\frac{dv}{2\pi(u-v)} \chi_{0}'(v) = 0\, ,
\eeq
with the solution
\beq\label{LO-R}
\chi_{0}'(u) = -\frac{2\rho}{u\sqrt{1-B^2/u^2}} \qquad \Leftrightarrow \qquad R_{0}(u) = \frac{\rho}{\sqrt{B^2-u^2}}\, .
\eeq
Straightforward integration yields the pseudo energy
\beq\label{chi0}
\chi_{0}(u) = -2\rho\log{\bigg[\frac{u+\sqrt{u^2-B^2}}{B}\bigg]}\, ,
\eeq
where the constant $C_{0} = 2\rho \log{(\tfrac{1}{2}B)}$ was fixed by demanding that $\chi_{0}$ vanishes at $\pm B$. This solution is the seed for the next steps and the full iteration can be taken at once by means of a suitable ansatz. Drawing inspiration from \cite{Volin:2009wr}, we write the general solution as an infinite series of terms, with increasing singularities at $u = \pm B$,
\beq\label{ansatz1}
R(u) = \sqrt{B}\sum_{n, m=0}^{\infty}\sum_{k = 0}^{n+m} \frac{c_{n, m, k}}{B^{m-n}(B^2-u^2)^{n+1/2}} (\frac{u}{B})^{[k]}\log^{k}{(\frac{B-u}{B+u})}\, ,
\eeq
where $[k] = k\,\, \textrm{mod}\,\,  2$ and where the coefficients $c_{n, m, k}$ are polynomials in $\log{B}$, up to an overall factor, see equation (\ref{cQ}) below. One easily verifies that the ansatz (\ref{ansatz1}) goes through equation (\ref{DR}) at every order in $1/B$. Not all coefficients are arbitrary though. The coefficients $c_{n, m, k\neq 0}$ for the logarithms are determined iteratively. For instance, at the next-to-leading order, the equation is fulfilled only if $c_{1, 0, 1} = -c_{0, 0, 0}/2\pi$. Also, the iteration cannot produce logarithmic enhancement of the leading square root behaviour (\ref{LO-R}), implying that $c_{0, m, k} = 0$ if $k\neq 0$. The remaining coefficients stay undetermined and are associated to homogeneous solutions. (The overall $\sqrt{B}$ in (\ref{ansatz1}) was put for later convenience.) Finally, note that the density $\rho$, which relates to the large $u$ behaviour, reads
\beq\label{rho-c}
\rho = \sqrt{B} \sum_{m=0}^{\infty} c_{0, m, 0}/B^m\, .
\eeq
The ansatz (\ref{ansatz1}) is identical to the one for the compact model if not for the support. Formally, one can get from one to the other by replacing $B- u \rightarrow u- B$ under the square root and inside the logs.

The ansatz (\ref{ansatz1}) solves the problem in generic terms in the domain connected to the boundary conditions at $u=\pm i\infty$. To fix its free parameters we must carry out a similar analysis close to the edge of the Fermi sea, at $u= B$. In particular, $\chi'$ should be regular at this point, since both the driving term and the kernel are smooth functions. This property is not built into the ansatz (\ref{ansatz1}), which shows an accumulation of singular behaviours at $u = B$. The coefficients must thus be adjusted such that their sum has a smooth discontinuity at $u=B$.

One can zoom on the neighbourhood of the branch point of the resolvent by using the variable $z = B-u$. Then, re-expanding (\ref{ansatz1}) at large $B$, with $z$ kept fixed,
\beq\label{DSL}
\begin{aligned}
R(B-z) &=  \sum_{n, m=0}^{\infty}\sum_{k = 0}^{n+m} \frac{c_{n, m, k} (1-z/B)^{[k]}}{B^{m}(2z(1-z/2B))^{n+1/2}} \log^{k}{(\frac{z}{2B-z})} \\
&= \frac{1}{\sqrt{2z}}\bigg[c_{0, 0, 0} + \frac{c_{1, 0, 0}+c_{1, 0, 1}\log{(z/2B)}}{z} + O(1/z^2)\bigg] + O(1/B)\, ,
\end{aligned}
\eeq
indicates that the series gets re-organised into a pertubative expansion with each coefficient being a non-trivial function of $z$, of which we only see the large $z$ tail here. The goal is now to determine all these functions for $z = O(1)$. To this end, it is convenient to introduce the Laplace transform
\beq\label{Laplace}
F(s) =  -\int\limits_{B}^{\infty}\frac{du}{2\pi}\, e^{-su} \chi'(u) = -s\int\limits_{B}^{\infty}\frac{du}{2\pi}\, e^{-su} \chi(u)\, ,
\eeq
which is a positive definite analytical function of $s$ for $s>0$. Since $\chi'$ is the discontinuity of the resolvent, see equation (\ref{R-disc}), we can also write the Laplace transform as a contour integral around $(B, \infty)$ or equivalently as
\beq
F(s) = \int\limits_{B-i\infty}^{B+i\infty}\frac{du}{2\pi i}\, e^{-su} R(u)\, .
\eeq
Notice also that the Laplace transform (\ref{Laplace}) determines the energy, which reads, in the low momentum approximation,
\beq\label{E-c}
E_{2d}/L \sim \frac{m}{4}\int\limits_{B}^{\infty}du\, e^{-\pi u/2} \chi(u)  = -mF(\frac{\pi}{2})\, .
\eeq
That one is negative, since $F$ is positive. Now, the behaviour (\ref{DSL}) translates into a non trivial statement about the expansion of $F(s)$ at large $B$, for $1/B \ll s \leqslant 1$. Namely,
\beq\label{FDSL}
\begin{aligned}
F(s) = e^{-Bs}\int\limits_{0-i\infty}^{0+i\infty}\frac{dz}{2\pi i}\, e^{sz} R(B-z) = \frac{e^{-Bs}}{\sqrt{2\pi s}} (F_{0}(s) + F_{1}(s)/B + \ldots) \, ,
\end{aligned}
\eeq
where $F_{0}(s) = c_{0, 0, 0} + O(s), F_{1}(s) = \#/s+ O(s^0),$ etc. We shall now proceed with the determination of these functions.

The most important piece of information comes from the analytical properties of $F(s)$. As a Laplace transform $F(s)$ is analytic for $\Re\textrm{e}\, s >0$. However, it must have a logarithmic branch point at $s=0$,
\beq
F(s) \sim \frac{\rho}{\pi}\log{(1/s)}\, ,
\eeq
in response to the the large $u$ behaviour of $\chi'$.
By contour manipulation, we can access to the discontinuity of $F(s)$ across $s<0$. First, note that rotating the contour of integration in (\ref{Laplace}) into the upper or lower half plane allows us to continue $F(s)$ to $\Im \textrm{m} \, s \ou 0$. Then, by taking the difference, we obtain an integral representation for the discontinuity,
\beq
\textrm{disc}\,  F(s) = F(s+i0)-F(s-i0) = \int \limits_{b-i\infty}^{b+i\infty}\frac{du}{2\pi}\,  e^{-su}\chi'(u)\, ,
\eeq
where $b$ is an arbitrary real number greater than $B$. The integral is zero for $s>0$, since then one can send $b\rightarrow \infty$. On the other hand, for $s<0$, we can evaluate the discontinuity by taking the inverse Laplace transform of the derivative of the equation (\ref{chi-eq}). Using
\beq
\int\limits_{0-i\infty}^{0+i\infty}\frac{du}{2\pi} \, e^{-su}\partial_{u}\epsilon(u) = 2i \cos{(\tfrac{1}{2}s)}\, , \qquad \int\limits_{0-i\infty}^{0+i\infty}\frac{du}{2\pi} \, e^{-su}\partial_{u}\mathcal{K}(u) = \frac{is}{\pi} \frac{\cos{(\tfrac{1}{2}s)}}{\sin{(\tfrac{1}{2}s)}}\cos{s}\, ,
\eeq
one obtains the relation
\beq\label{disc}
\textrm{disc}\,  F(s) = -2i\cos{(\tfrac{1}{2}s)}\bigg[\frac{m}{4\sqrt{2}} - \frac{s\cos{s}}{\sin{(\tfrac{1}{2}s)}} \times \int\limits_{-B}^{B}\frac{du}{2\pi}\, e^{-su} \chi(u)\bigg]\, ,
\eeq
where to make the units apparent we introduced $m/4\sqrt{2} = 1$ in front of the driving term. On the RHS we find the Laplace transform of $\chi$ on the inner support, which is compact. The latter function is strictly positive for $s$ real, since $\chi$ is positive on this interval. Therefore, we readily derive from (\ref{disc}) that $\textrm{disc}\, F(s)$ has simple poles at $s = -2\pi k$ for $k\in \mathbb{N}$ and nowhere else. We also observe that $\textrm{disc}\, F(s)$ is completely determined at $s = s_{n} = -\pi (n-\tfrac{1}{2})$ by the driving term.  Since, according to (\ref{FDSL}), $\textrm{disc}\, F(s) \propto e^{-Bs}$ at large $B$, the driving term appears exponentially small compared to the rest. Therefore, up to negligible corrections, its sole effect is to set the discontinuity at $s = s_{1} = -\tfrac{1}{2}\pi$ and one can assume that the discontinuity vanishes at $s=s_{n>1}$.

These analytical properties of $F(s)$ together with the requirement that it should admit an expansion in powers of $1/s$ at large $s$, for the regularity of $\chi'$ at $u = B$, fix the shapes of the functions in (\ref{FDSL}). The same problem was addressed in \cite{Volin:2009wr} for the compact model and its general solution was given in the form
\beq\label{ansatz2}
F(s) = e^{-Bs}e^{\frac{s}{2\pi}\log{(2s/\pi e)}}\frac{A\sqrt{s}\, \Gamma(\frac{s}{2\pi})}{4\sqrt{2}\Gamma(\tfrac{1}{2}+\tfrac{s}{\pi})} (\frac{1}{s+\tfrac{\pi}{2}}+Q(s))\, ,
\eeq
where $Q(s) = \sum_{n, m\geqslant 0}Q_{n, m}/(B^{m+n+1}s^{n+1})$ represents a general zero mode solution. Comparing the discontinuity of (\ref{disc}) at $s=-\pi/2$ determines the overall scale,
\beq\label{A}
A = \frac{me^{-\frac{\pi}{2}B-\frac{1}{4}}}{\sqrt{\pi}}\Gamma(\tfrac{5}{4})\, .
\eeq

We can now reap the fruits of our labor (actually Volin's labor): confronting (\ref{ansatz1}) and (\ref{ansatz2}) unequivocally fixes all the free coefficients, with the first few of them given explicitly by
\beq\label{cQ}
c_{0, 0, 0} = A\, , \qquad c_{0, 1, 0} = \frac{A}{4\pi}(3+\log{(\tfrac{1}{4}\pi B)})\,  , \qquad Q_{0, 0} = -\frac{1}{4\pi}\, .
\eeq
Then, using (\ref{rho-c}) and (\ref{E-c}), we obtain energy and charge density, order by order in $1/B$,
\beq\label{final-E-rho}
\begin{aligned}
\rho  = A\sqrt{B}(1+\frac{3+\log{(\tfrac{1}{4}\pi B)}}{4\pi B} + \ldots)\, ,\qquad E_{2d}/L = -\frac{A^2}{2}(1-\frac{1}{2\pi B} +\ldots)\, .
\end{aligned}
\eeq
Note in particular that both are exponentially small, since $A\propto e^{-\pi B/2}$. Finally, after eliminating $B\sim \frac{2}{\pi}\log{(m/\rho)}$, we arrive at
\beq\label{E-int}
E_{2d}/(\pi L\rho^2) = -\frac{1}{4\log{(m c/\rho)}} + \frac{3\log{\log^2{(mc/\rho)}}+4}{32\log^2{(mc/\rho)}} + \ldots\,.
\eeq
where $c = (2e)^{1/4}\Gamma(\tfrac{5}{4})/\pi$. This formula fits nicely the numerical result shown in figure \ref{numerics} and it agrees with the sigma model prediction (\ref{E-pert}) provided one identifies the mass scales as in (\ref{L-to-m}). 

The agreement is guaranteed by the perturbative relation to the $O(6)$ model. Namely, one can formally obtain the all order formulae for energy and charge density in the $O(6)$ model by flipping the sign of $B$ in (\ref{final-E-rho}), ignoring the logs. (One must also use $m\rightarrow -im$ to keep $\rho$ real.) Indeed, applying this rule at given $z$ in (\ref{DSL}), disregarding the minus signs appearing in the arguments of the logs, reproduces the formula of \cite{Volin:2009wr} for the distribution density of the compact model in the region $u = -B-z$ with $z = O(1)$. The rest of the analysis, the matching with the Laplace transform (\ref{ansatz2}), is the same in both cases. For the record, the precise map among coefficients is
\beq
\begin{aligned}
&A' = \sqrt{2\pi}\, A\, |_{B\rightarrow -\frac{2B'}{\pi}} \, , \qquad c_{n, m, k}'/A' = 2(-1)^{m} c_{n, m, k}/A\, |_{\log{B}\rightarrow \log{\frac{2B'}{\pi}}}\, , \\
&Q'_{n, m} = (-\tfrac{1}{2})^{m+n+1}\pi^{m+1}Q_{n, m}\, |_{\log{B}\rightarrow \log{\frac{2B'}{\pi}}}\, ,
\end{aligned}
\eeq
where the primed quantities refer to the expressions for the $O(6)$ model \cite{Volin:2009wr} and where the numerical factors result from the $u$ to $\theta$ conversion.

In the end, since $B$ behaves like the radius of the target space, the $B\rightarrow -B$ rule mimics the field theory recipe for connecting the sphere and the hyperboloid perturbatively. Note however that eliminating $B$ yields to formula (\ref{E-int}) in any case. The difference lies then entirely in the domain of validity of this formula, that is high densities, $\rho\gg m$, for the sphere and low densities, $\rho \ll m$, for the hyperboloid.

\section{Spinning the wheels}\label{App2}

In this appendix we discuss a simple class of excited states obtained by adding spin to the operator,
\beq
\mathcal{O}\sim \textrm{tr}\, \partial^{M}\phi_{1}^{L}\ ,
\eeq
with all the derivatives pointing in the same light-like direction, $\partial = n^{\mu}\partial_{\mu}$ with $n^2 = 0$. We expect these operators to correspond to states with transverse excitations $\sim (X_{1}+iX_{2})^{M}$ in the sigma model. Below we obtain Bethe ansatz equations for them and, with their help, test the correspondence with the sigma model energy levels $E_{2d}(L, \Delta, M)$.

Adding derivatives to the operator brings an extra layer of difficulty since one must diagonalize a mixing problem. In the one-loop planar gauge theories this mixing results from gluon exchange and leads to the appearance of the $\mathfrak{sl}(2)$ XXX Hamiltonian. There are no gluons, nor any short-ranged spin chain interactions, in the fishnet theory in this sub-sector of operators. So, here again, the 2d dynamics originates from virtual scalar fields traveling all around the operators. Nevertheless, the eigenstates turn out be the same, to leading order at weak coupling, as those for the Heisenberg spin chain. The point is that the twist that connects the fishnet theory to $\mathcal{N}=4$ SYM does not affect the Bethe ansatz equations in the $\mathfrak{sl}(2)$ subsector. Hence, conformal primaries are described by Bethe states, that are linear superpositions of spin waves of derivatives on top of the BMN vacuum. They are generated by the action of the spin-chain creation operator $B(v)$,
\beq
B(v_{1})\ldots B(v_{M}) \textrm{tr}\, \phi_{1}^{L}\, ,
\eeq
and are parameterized by the Bethe rapidities $\textbf{v} = \{v_{1}, \ldots , v_{M}\}$. For a periodic spin chain, the rapidities are quantized by the Bethe ansatz equations,
\beq\label{oBAE}
1 = \big(\frac{v_{k}-\tfrac{i}{2}}{v_{k}+\tfrac{i}{2}}\big)^{L} e^{i\Phi(v_{k})} \prod_{j\neq k}^{S}\frac{v_{k}-v_{j}-i}{v_{k}-v_{j}+i}\, ,
\eeq
where $\Phi =0$ to leading order at weak coupling. The scaling dimension of the corresponding operator, in the absence of short-ranged interactions, is given to leading order at weak coupling by the L\"uscher formula \cite{Bajnok:2008bm,Bajnok:2008qj}. Once reduced to the fishnet theory, one finds
\beq\label{L1}
\Delta = L+M -2g^{2L}\sum_{a\geqslant 1}\int\limits_{-\infty}^{\infty}\frac{du}{2\pi} e^{-L\epsilon_{a}(u)} \prod_{j=1}^{M}S_{a*}(u, v_{j})T_{a}(u)\dot{T}_{a}(u) + O(g^{4L})\, .
\eeq
where $S_{a*}(u, v)$ is the diagonal part of the scattering matrix between a mirror scalar magnon and a derivative, and where $T_{a} = \dot{T}_{a}$ is a rational transfer matrix in the spin $\tfrac{1}{2}(a-1)$ representation of $SU(2)$. The gas of wheels also induces finite size corrections to the Bethe roots, represented in (\ref{oBAE}) by the additional phase shift $\Phi = O(g^{2L})$. The leading-order formula can also be obtained from \cite{Bajnok:2008bm,Bajnok:2008qj} and reads
\beq\label{L2}
\Phi(v) = g^{2L}\sum_{a\geqslant 1}\int \frac{du}{2\pi} e^{-L\epsilon_{a}(u)} \textrm{tr}\,\{ \partial_{u}\mathcal{S}_{a*}(u, v_{k})\prod_{j\neq k}^{M}\mathcal{S}_{a*}(u, v_{j})\} + O(g^{4L})\, ,
\eeq
where the trace is taken over the $O(4)$ indices of the S-matrix $\mathcal{S}_{1*}(u, v)$ between a mirror magnon and a derivative.

The all order formula for the scaling dimension takes the same form as before, though the Y functions are now shifted by extra source terms which accommodate for the $\textbf{v}$-dependent part of the integrand in (\ref{L1}). The general formula for the phase shift $\Phi$ is not known to us, except for its s-wave component. In the latter case the scattering is abelian, $\mathcal{S}_{1*}\rightarrow S_{1*}$, and there is no need for the trace in (\ref{L2}). By the same token, there are no transfer matrices in (\ref{L1}), $T_{1} = \dot{T}_{1} = 1$. In these circumstances, one can immediately write down the all order formula
\beq\label{Phi-s}
\Phi(v) = \int\limits_{-\infty}^{\infty} \frac{du}{2\pi}\partial_{u}\log{S_{1*}(u, v)} \log{(1+Y_{1}(u))}\, ,
\eeq
which is valid up to the contributions of the heavier magnons, $a>1$. The only ingredient is the S matrix%
\footnote{Derivatives and mirror magnons live in different kinematics. This is why (\ref{Swd}) does not look unitary.}
\beq\label{Swd}
S_{1*}(u, \textbf{v}) = \prod_{j=1}^{M} S_{1*}(u, v_{j}) = \frac{Q(\tfrac{i}{2})Q(-\tfrac{i}{2})}{Q(u+i)Q(u-i)}\, ,
\eeq
with $Q(u) = \prod_{j=1}^{M}(u-v_{j})$ the Baxter polynomial, and the same factor should be added to the RHS of the $Y_{1}$ TBA equation, which is shifted by
\beq\label{extra}
\delta \log{Y_{1}} = \log{S_{1*}(u, \textbf{v})}\, .
\eeq
Note that this shift does not alter the asymptotic behaviour of the Y function, which reads
\beq
\log{Y_{1}} \sim -2(L+M+\gamma)\log{u} = -2\Delta \log{u}\, ,
\eeq
with the anomalous dimension $\gamma = - \sum_{a}\int dp\log{(1+Y_{a})}/(2\pi)$. Indeed, the extra source term (\ref{extra}) only brings the ``$M$'' in this expression, which is the total bare dimension for the derivatives in the state.

\begin{figure}
\begin{center}
\includegraphics[scale=0.4]{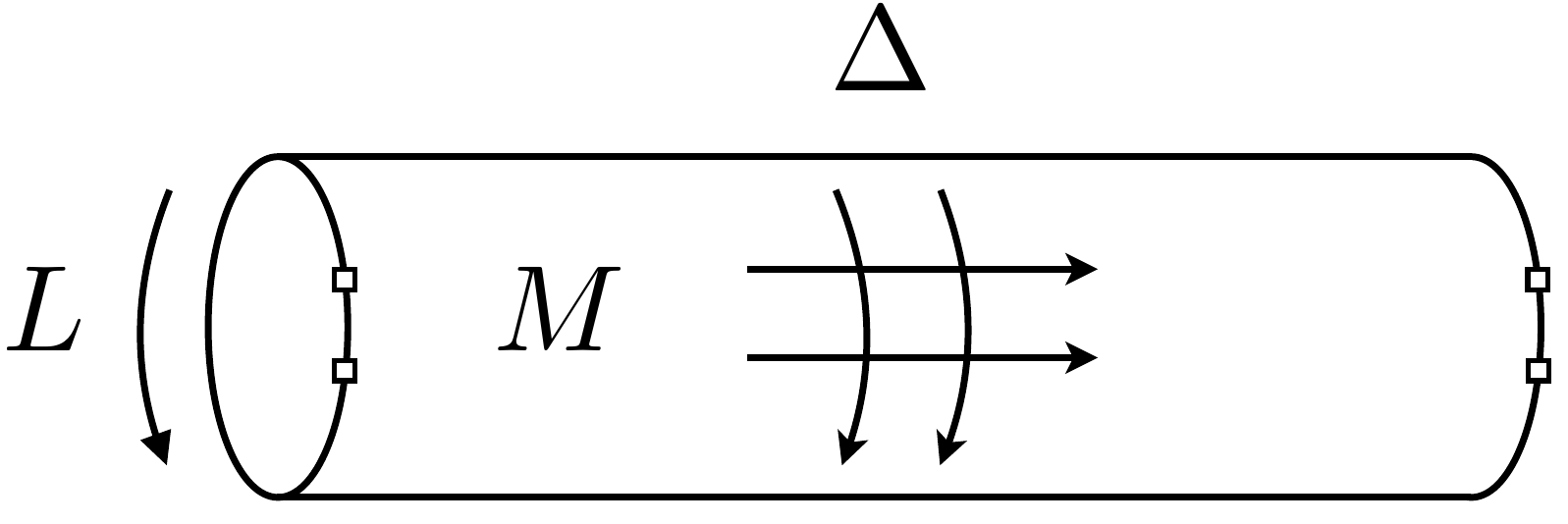}
\end{center}
%\vspace{-8cm}
\caption{The longitudinal excitations $\sim (X_{0}+iX_{d+1})^{\Delta}$ in the thermal bath cross the cylinder and scatter with the transverse excitations $\sim (X_{1}+iX_{2})^{M}$ in the state. The amplitude for this process is controlled by $\sigma_{2}(u+i, v)$, where the shift of the first argument implements the rotation to the crossed channel, $\theta \rightarrow \theta +i\pi/2$.}\label{cylinder}
\end{figure}

We can now proceed with the dualization of the equations. We begin with the equation for the condensate. We just need to act with $(1-K_{O(6)}*)$ on the extra source term (\ref{extra}). Straightforward algebra gives
\beq
(1 - K_{O(6)}*)\log{S_{1*}(u, \textbf{v})} = -\sum_{k=1}^{M} \log{\sigma_{2}(u+i-v_{k})}\, ,
\eeq
where
\beq\label{sigma2}
\sigma_{2}(u) = \frac{u}{u-i}S_{O(6)}(u)\, ,
\eeq
and thus the dual TBA equation is
\beq\label{Yd}
\log{Y_{1}} = LE -\log{\sigma_{2}(u-\textbf{v}+i)} - K_{O(6)}*\log{(1+1/Y_{1})} + \ldots\, ,
\eeq
where the dots stand for the couplings to the higher modes, see equation (\ref{main}). The amplitude (\ref{sigma2}) has a nice interpretation: it is the transmission amplitude for the scattering of two orthogonal complex scalar fields, $X_{0}+iX_{d+1}$ and $X_{1}+iX_{2}$, in the $O(6)$ model. Hence, the Y function differs in that the thermal excitations it describes pick an extra contribution owing to their scattering with the derivatives in the state, as shown in figure \ref{cylinder}.

A similar analysis can be done for the energy of the state which becomes
\beq
L \log{g^2/g_{cr}^2}  = E_{2d} = \sum_{i=1}^{M}E(v_{i})-\int\limits_{-\infty}^{\infty}\frac{du}{2\pi}\, P'(u)\log{(1+1/Y_{1}(u))}\, ,
\eeq
using
\beq
\int\limits_{-\infty}^{\infty}\frac{du}{2\pi}\, P'(u) \log{\frac{Q(u+i)Q(u-i)}{Q(\tfrac{i}{2})Q(-\tfrac{i}{2})}} = \sum_{i=1}^{M} E(v_{i})\, .
\eeq
Nicely $E_{2d}$ is the sum of the mechanical energies of the transverse excitations and of the tachyon energy. Furthermore, the transverse excitations contribute positively to the total energy, in agreement with the expectation based on the AdS model.

Equation (\ref{Yd}) captures the pseudo-energy shift induced by the derivatives. The reciprocal action can be found by dualizing equation (\ref{oBAE}). One first rewrites (\ref{Phi-s}) as
\beq
\Phi(v) =  \int \frac{du}{2\pi}\partial_{u}\log{S_{1*}(u, v)} \log{Y_{1}(u)} + \ldots \, ,
\eeq
and then plug (\ref{Yd}) in its RHS. The identities
\beq
\begin{aligned}
&\int \frac{du}{2\pi}\partial_{u}\log{S_{1*}(u, v)} E(u) = i\log\big(\frac{v-i/2}{v+i/2}\big) + P(v)\, , \\
&\int \frac{du}{2\pi}\partial_{u}\log{S_{1*}(u, v)} \log{\sigma_{2}(u-w+i)} = -i\log\big(\frac{v-w-i}{v-w+i}\big) +i\log{S_{O(6)}(v-w)}\, ,
\end{aligned}
\eeq
show that the spin chain interactions are erased and replaced by the sigma model ones. The net result is
\beq\label{n-BAE}
1 = e^{iP(v_{k})L+i\Psi(v_{k})} \prod_{j\neq k}^{M}S_{O(6)}(v_{k}-v_{j})\, ,
\eeq
where 
\beq
\Psi(v) = -\int\frac{du}{2\pi}\partial_{u}\log{\sigma_{2}(u+i-v)}\log{(1+1/Y_{1}(u))} + \ldots\, ,
\eeq
and with the dots representing possible couplings to the Y functions that we discarded.

In the following we set the tachyon component to zero, $\Psi = 0$. Our goal is to compare the BAEs (\ref{n-BAE}) with the similar equations obtained in the compact sigma model. It does not hurt being slightly more general and consider the $d$ dimensional version of the problem. In logarithmic form, after going to hyperbolic rapidities, we get to solve the equations
\beq
-LP_{k} = 2\pi n_{k} -\sum_{j\neq k}^{M}i\log{S_{O(d+2)}(\theta_{k}-\theta_{j})}\, .
\eeq
They can only be trusted at large $L$, which means small momenta, $P_{k} \sim \mp \frac{m}{2} e^{- |\theta_{k}|}\sim 0$. The roots are thus pushed towards $\theta = \pm \infty$ and split into left and right movers, $\theta^{\pm}_{k}$, depending on the sign of their mode number $n_{k}$. Shifting the roots according to their chirality,
\beq\label{shifts}
\theta^{\pm}_{k} \rightarrow \theta^{\pm}_{k} \pm \xi\, ,
\eeq
where $\xi = \log{(\frac{1}{2}mL)}\gg 1$, yields the equivalent equations
\beq\label{masslessBA}
e^{\mp \theta^{\pm}_{k}} = 2\pi n^{\pm}_{k} \mp  i\sum_{j\neq k}^{M_{\pm}} \log{S_{O(d+2)}(\theta^{\pm}_{k}-\theta^{\pm}_{j})} \mp i\sum_{j}^{M_{\mp}} \log{S_{O(d+2)}(\theta^{\pm}_{k}-\theta^{\mp}_{j}\pm 2\xi)}\, ,
\eeq
where $n^{\pm}_{k}$ are positive integers, regardless of the chirality.%
\footnote{Note also that there cannot be vanishing mode numbers, $n_{k}\neq 0$.}
In this form it is clear that left and right movers decouple at large $L$, since \cite{Zamolodchikov:1977nu,Zamolodchikov:1978xm}
\beq\label{asy-S}
i\log{S_{O(d+2)}(2\xi)} \sim \frac{\pi}{d\xi} \sim \frac{e^{2}}{2} \ll 1\, ,
\eeq
where $e^2\sim 2\pi/(d\log{(mL)})$ is the coupling at distance $L$. Hence, the perturbative expansion is controlled by the strength $\sim 1/\xi$ of the left-right interactions. The leading order ``scale invariant'' or kink solutions are obtained by discarding these interactions. Since the remaining interactions, among particles of same chirality, are of order $O(1)$, the kink solutions can only be found numerically, for generic distributions of mode numbers. Fortunately we will not need to know them explicitly.

What is important is that the massless Bethe ansatz equations (\ref{masslessBA}) also apply to the compact model, with the twist that they hold for \textit{small} length, $L\ll 1$. Hence, the equations stay the same but $\xi = - \log{(2/mL)}$ is now large and \textit{negative}. (Also, left and right movers exchange their roles, $\theta^{\pm}_{k} \sim \mp \log{(2/mL)} \sim \mp \infty$, when going from compact to non-compact.) We could also bring the equations to a more standard form by performing $\xi \rightarrow -\xi$, which has the same effect as flipping the sign of the coupling.

Lastly, let us discuss the direct comparison with the sigma model energy levels. A well known trick allows us to evaluate the first two perturbative terms in the energy, without actually solving the equations. Shifting the roots as in (\ref{shifts}) gives
\beq
E_{2d} \simeq \frac{1}{L}\sum_{j=1}^{M_{+}} e^{-\theta^{+}_{j}} +\frac{1}{L}\sum_{j=1}^{M_{-}} e^{\theta^{-}_{j}}\, ,
\eeq
and summing over the RHS of (\ref{masslessBA}) yields
\beq
E_{2d} = \frac{2\pi}{L}(N_{+}+N_{-}) -\frac{2i}{L}\sum_{j, k}\log{S_{O(d+2)}(\theta_{j}^{+}-\theta_{k}^{-}+2\xi)}\, ,
\eeq
where we used that $\log{S(\theta)} = -\log{S(-\theta)}$ and where $N_{\pm} = \sum_{j=1}^{M_{\pm }}n^{\pm}_{j}$ are the levels for left and right movers. Finally, using (\ref{asy-S}), one concludes that
\beq\label{BAresult}
E_{2d} \simeq \frac{2\pi}{L}(N_{+}+N_{-}) -\frac{e^2}{L} M_{+}M_{-}\, , 
\eeq
with $e^2\sim 2\pi/(d\log{(mL)})$.

Energy levels on a cylinder can be computed in the CFT limit using the operator-state correspondence. In our case this is applicable in the IR, at large $L$. The vertex operators of interest take the form
\beq
V_{M} \sim \partial_{+}^{N_{+}}(X_{1}+iX_{2})^{M_{+}}\partial^{N_{-}}_{-}(X_{1}+iX_{2})^{M_{-}} \, ,
\eeq
where $\partial_{\pm}$ are worldsheet light-cone derivatives. The 2d scaling dimensions $\delta$ are known through one loop in the sphere sigma model and converting them to the $AdS_{d+1}$ space yields \cite{Wegner,Tseytlin:2003ac}
\beq
\delta = N_{+}+N_{-}  - \frac{e^2(M_{+}M_{-}+\frac{1}{2}M(d+1))}{2\pi} + O(e^4)\, .
\eeq
They correspond to the energy levels, up to an overall factor $2\pi/L$ and up to the Casimir energy.

Comparison with the BA expression (\ref{BAresult}) gives a match modulo the correction $\sim -e^2 (d+1) M/(2L)$. The latter contribution stems from the renormalization of the fundamental fields in the sigma model. It relates to the small volume ``mass gap'' which entirely comes from the finite size corrections in the compact case \cite{Balog:2003yr,Balog:2005yz}. Therefore, it is not surprising to find that it evades the BAEs here as well. Nonetheless, it is interesting to find that the BAEs capture the rest correctly. In the sigma model analysis, this part is the most interesting one and comes from the exchange contribution in the terminology of \cite{Wegner,Tseytlin:2003ac}. It would be nice to see if this correspondence holds for generic states-operators, including those carrying non zero $\Delta$.

Finally, let us mention that there is a cleaner way of testing the BAEs, by going to the semi-classical regime of large quantum numbers, $e^2 M_{\pm} = O(1)$. A systematic analysis of this regime, like the one carried out in \cite{Gromov:2006dh} for the sphere, could reveal how much of the Hilbert space of the sigma model is realized in the fishnet theory.

\section{3d fishnet graphs}\label{App3}

In this appendix we discuss the thermodynamical limit of the triangular fishnet graphs in 3d. These graphs lie at the end point of a suitable twisting of the weakly coupled ABJM theory \cite{Caetano:2016ydc}. The corresponding 3d fishnet theory  admits the Lagrangian
\beq
\mathcal{L}_{3d} = N\, \textrm{tr}\, \bigg[\sum_{i=1,2,3}\partial_{\mu}\phi_{i}\partial_{\mu}\phi_{i}^*+(4\pi g)^2\,\phi_{1}\phi_{2}^{*}\phi_{3}\phi_{1}^{*}\phi_{2}\phi_{3}^{*}\bigg]\, ,
\eeq
where every complex scalar field carries $U(N)\times U(N)$ bi-fundamental indices, and the trace is taken over these indices. Since the gauge fields of the ABJM theory are entirely decoupled at the end of the twisting procedure, the latter group is merely a flavour group. The planar limit corresponds to sending $N\rightarrow \infty$ at fixed $g^2$ and the theory is conformal for any $g^2$. In particular, no double-trace interactions are needed here \cite{Mamroud:2017uyz}.

The simplest single-trace operator is the BMN vacuum operator
\beq
\textrm{tr}\, (\phi_{1}\phi_{2}^{*})^{L}\, ,
\eeq
with engineering dimension $L$. Its anomalous dimension in the planar limit is induced by virtual particles of the field $\phi_{3}$ that loop around the operator, as displayed in \cite{Caetano:2016ydc}. In the integrability framework, these particles map to mirror magnons and the scaling dimension computes the free energy of the gas of magnons at temperature $1/L$. The mechanical energies carried by the mirror magnons take the same form (\ref{eps-a}) as in the 4d theory, but their momentum, at a rapidity $u$, is half the one in the 4d theory, $p = u$. In the thermodynamical limit $L\rightarrow \infty$ the s-wave magnons, with $a=1$, will condense for $g^2>1/4$ and fill a Fermi sea. The scaling dimension scales then thermodynamically
\beq\label{D3d}
\Delta/L = 1 -\int\limits_{-B}^{B}\frac{du}{2\pi}\, \chi(u)\, ,
\eeq
where the pseudo energy $\chi$ solves the Bethe ansatz equation (\ref{chi-eq}), with the boundary condition $\chi(u = \pm B) = 0$, the 3d scattering kernel
\beq\label{calK-3d}
\mathcal{K}(u) = \psi(2+iu)+\psi(2-iu)\, ,
\eeq
the constant
\beq\label{effc}
C = \log{g^2} - \frac{1}{2}\int\limits_{-B}^{B}\frac{du}{2\pi}\, k(u)\chi(u)\, ,
\eeq 
and with $k(u)$ as in (\ref{Ku}). These identities follow directly from the relation between S-matrices for mirror magnons in the 3d and 4d theory \cite{Gromov:2008qe,Ahn:2008aa}
\beq
S(u, v)|_{3d} = -\sqrt{\frac{u-v-i}{u-v+i}S(u, v)}|_{4d}\, .
\eeq

In the limit where all the energy levels are filled, that is for $B=\infty$, the solution to the integral equation is obtained by going to Fourier space and reads
\beq\label{Chi-cr-3d}
\chi_{cr}(u) = \int\limits_{-\infty}^{\infty}dt\, e^{iu t}\frac{\sinh{t}}{t\cosh{(\tfrac{3t}{2})}} = \log{\left[\frac{\cosh{(\tfrac{\pi u}{3})}+\tfrac{\sqrt{3}}{2}}{\cosh{(\tfrac{\pi u}{3})}-\tfrac{\sqrt{3}}{2}}\right]}\, .
\eeq
At this point, the scaling dimension (\ref{D3d}) and effective chemical potential (\ref{effc}) both vanish. The latter condition determines the critical coupling,
\beq
\log{g^{2}_{cr}} = \int\limits_{-\infty}^{\infty}\frac{du}{4\pi}\, k(u) \chi_{cr}(u) = 2\int\limits_{0}^{\infty}\frac{dt}{t}(e^{-t}-\frac{(e^{t}+1)^2}{2(e^{3t}+1)}) = 3\log{\left[\frac{12\sqrt{\pi}\Gamma(\tfrac{2}{3})}{\Gamma(\tfrac{1}{6})}\right]}-\log{(4\pi)^2}\, ,
\eeq
or, numerically, $g_{cr} = 0.936...$. It agrees with Zamolodchikov's prediction \cite{Zamolodchikov:1980mb}.

The neighbourhood of the critical point is described by the dual equation (\ref{dual-chi-eq}), sourced by the dual energy $E = \chi_{cr}$ coming from (\ref{Chi-cr-3d}). The latter is seen to decay exponentially fast at large rapidity,
\beq\label{E3d}
E = \frac{m}{2} e^{-|\theta|}\, ,
\eeq
where $m = 4\sqrt{3}$ and with $\theta = \pi u/3$. Hence, here again, the dual excitations are gapless. The dual scattering kernel is obtained by solving (\ref{dual-K}) for $\mathcal{K}$ as given in (\ref{calK-3d}) and reads
\beq\label{KO5}
K(u) = 2\int\limits_{0}^{\infty}dt\, \frac{e^{t}+1}{e^{3t}+1}\cos{(u t)} = \frac{\partial}{i\partial u}\log{\frac{\Gamma(+\tfrac{iu}{6})\Gamma(\tfrac{1}{2}-\tfrac{iu}{6})\Gamma(\tfrac{5}{6}+\tfrac{iu}{6})\Gamma(\tfrac{1}{3}-\tfrac{iu}{6})}{\Gamma(-\tfrac{iu}{6})\Gamma(\tfrac{1}{2}+\tfrac{iu}{6})\Gamma(\tfrac{5}{6}-\tfrac{iu}{6})\Gamma(\tfrac{1}{3}+\tfrac{iu}{6})}}\, .
\eeq
It is the same as the scattering kernel for identical particles in the non-linear $O(5)$ sigma model \cite{Zamolodchikov:1977nu,Zamolodchikov:1978xm}.

Expressions (\ref{E3d}) and (\ref{KO5}) hint at a connection between the planar 3d fishnets and the $AdS_{4}$ sigma model.

There is a little twist, compared to what we had in 4d, which relates to the discrete, triangular, nature of the 3d lattice. Namely, the coupling constant of the 3d fishnet theory is not exactly the sigma model energy. The ``marginality condition'' still takes the form
\beq\label{MC3d}
\log{g^2} = \log{g^2_{cr}} + \int\limits_{u^2\geqslant B^2}\frac{du}{2\pi}\, \p_{cr}(u)\chi(u)\, ,
\eeq
but
\beq\label{pcr3d}
\p_{cr}(u) = -\frac{1}{2}(1-K*)k = \frac{1}{2i}\partial_{u}E(u+i) + \frac{1}{2i}\partial_{u}E(u+2i)\, ,
\eeq
is not identical to the derivative of the dual momentum $P(u) = -iE(u+3i/2)$. Instead, once written in terms of the hyperbolic rapidity $\theta$, the imaginary shifts in the RHS of (\ref{pcr3d}) are seen to implement rotations by $\pi/3$ and $2\pi/3$, respectively. These are the angles that characterize the triangular fishnets. However, the distinction is small at low energy,
\beq
\p_{cr}(u) \sim \frac{\sqrt{3}}{2} P'(u)\, ,
\eeq
after disregarding irrelevant, exponentially small, corrections.

\bibliography{biblio}

\providecommand{\href}[2]{#2}\begingroup\raggedright\begin{thebibliography}{10}

\bibitem{tHooft:1973alw}
G.~'t~Hooft, \emph{{A Planar Diagram Theory for Strong Interactions}},
  \href{http://dx.doi.org/10.1016/0550-3213(74)90154-0}{\emph{Nucl. Phys.} {\bf
  B72} (1974) 461}.

\bibitem{Aharony:1999ti}
O.~Aharony, S.~S. Gubser, J.~M. Maldacena, H.~Ooguri and Y.~Oz, \emph{{Large N
  field theories, string theory and gravity}},
  \href{http://dx.doi.org/10.1016/S0370-1573(99)00083-6}{\emph{Phys. Rept.}
  {\bf 323} (2000) 183--386}, [\href{https://arxiv.org/abs/hep-th/9905111}{{\tt
  hep-th/9905111}}].

\bibitem{Maldacena:1997re}
J.~M. Maldacena, \emph{{The Large N limit of superconformal field theories and
  supergravity}}, \href{http://dx.doi.org/10.1023/A:1026654312961}{\emph{Int.
  J. Theor. Phys.} {\bf 38} (1999) 1113--1133},
  [\href{https://arxiv.org/abs/hep-th/9711200}{{\tt hep-th/9711200}}].

\bibitem{Beisert:2010jr}
N.~Beisert et~al., \emph{{Review of AdS/CFT Integrability: An Overview}},
  \href{http://dx.doi.org/10.1007/s11005-011-0529-2}{\emph{Lett. Math. Phys.}
  {\bf 99} (2012) 3--32}, [\href{https://arxiv.org/abs/1012.3982}{{\tt
  1012.3982}}].

\bibitem{Basso:2013vsa}
B.~Basso, A.~Sever and P.~Vieira, \emph{{Spacetime and Flux Tube S-Matrices at
  Finite Coupling for N=4 Supersymmetric Yang-Mills Theory}},
  \href{http://dx.doi.org/10.1103/PhysRevLett.111.091602}{\emph{Phys. Rev.
  Lett.} {\bf 111} (2013) 091602}, [\href{https://arxiv.org/abs/1303.1396}{{\tt
  1303.1396}}].

\bibitem{Basso:2015zoa}
B.~Basso, S.~Komatsu and P.~Vieira, \emph{{Structure Constants and Integrable
  Bootstrap in Planar N=4 SYM Theory}},
  \href{https://arxiv.org/abs/1505.06745}{{\tt 1505.06745}}.

\bibitem{Fleury:2016ykk}
T.~Fleury and S.~Komatsu, \emph{{Hexagonalization of Correlation Functions}},
  \href{http://dx.doi.org/10.1007/JHEP01(2017)130}{\emph{JHEP} {\bf 01} (2017)
  130}, [\href{https://arxiv.org/abs/1611.05577}{{\tt 1611.05577}}].

\bibitem{Eden:2016xvg}
B.~Eden and A.~Sfondrini, \emph{{Tessellating cushions: four-point functions in
  $\mathcal{N} $ = 4 SYM}},
  \href{http://dx.doi.org/10.1007/JHEP10(2017)098}{\emph{JHEP} {\bf 10} (2017)
  098}, [\href{https://arxiv.org/abs/1611.05436}{{\tt 1611.05436}}].

\bibitem{Bargheer:2017nne}
T.~Bargheer, J.~Caetano, T.~Fleury, S.~Komatsu and P.~Vieira, \emph{{Handling
  Handles I: Nonplanar Integrability}},
  \href{https://arxiv.org/abs/1711.05326}{{\tt 1711.05326}}.

\bibitem{Eden:2017ozn}
B.~Eden, Y.~Jiang, D.~le~Plat and A.~Sfondrini, \emph{{Colour-dressed hexagon
  tessellations for correlation functions and non-planar corrections}},
  \href{http://dx.doi.org/10.1007/JHEP02(2018)170}{\emph{JHEP} {\bf 02} (2018)
  170}, [\href{https://arxiv.org/abs/1710.10212}{{\tt 1710.10212}}].

\bibitem{Ben-Israel:2018ckc}
R.~Ben-Israel, A.~G. Tumanov and A.~Sever, \emph{{Scattering Amplitudes --
  Wilson Loops Duality for the First Non-planar Correction}},
  \href{https://arxiv.org/abs/1802.09395}{{\tt 1802.09395}}.

\bibitem{Gurdogan:2015csr}
O.~Gurdogan and V.~Kazakov, \emph{{New Integrable 4D Quantum Field Theories
  from Strongly Deformed Planar $\mathcal N = $ 4 Supersymmetric Yang-Mills
  Theory}}, \href{http://dx.doi.org/10.1103/PhysRevLett.117.201602,
  10.1103/PhysRevLett.117.259903}{\emph{Phys. Rev. Lett.} {\bf 117} (2016)
  201602}, [\href{https://arxiv.org/abs/1512.06704}{{\tt 1512.06704}}].

\bibitem{Zamolodchikov:1980mb}
A.~B. Zamolodchikov, \emph{{Fishnet Diagrams as a Completely Integrable
  System}}, \href{http://dx.doi.org/10.1016/0370-2693(80)90547-X}{\emph{Phys.
  Lett.} {\bf 97B} (1980) 63--66}.

\bibitem{Caetano:2016ydc}
J.~Caetano, O.~Gurdogan and V.~Kazakov, \emph{{Chiral limit of $ \mathcal{N} $
  = 4 SYM and ABJM and integrable Feynman graphs}},
  \href{http://dx.doi.org/10.1007/JHEP03(2018)077}{\emph{JHEP} {\bf 03} (2018)
  077}, [\href{https://arxiv.org/abs/1612.05895}{{\tt 1612.05895}}].

\bibitem{Grabner:2017pgm}
D.~Grabner, N.~Gromov, V.~Kazakov and G.~Korchemsky, \emph{{Strongly
  $\gamma$-deformed N=4 SYM as an integrable CFT}},
  \href{https://arxiv.org/abs/1711.04786}{{\tt 1711.04786}}.

\bibitem{Kazakov:2018qbr}
V.~Kazakov and E.~Olivucci, \emph{{Bi-scalar integrable CFT at any dimension}},
   \href{https://arxiv.org/abs/1801.09844}{{\tt 1801.09844}}.

\bibitem{Kazakov:2018ugh}
V.~Kazakov, \emph{{Quantum Spectral Curve of $\gamma$-twisted ${\cal N}=4$ SYM
  theory and fishnet CFT}},  \href{https://arxiv.org/abs/1802.02160}{{\tt
  1802.02160}}.

\bibitem{Sieg:2016vap}
C.~Sieg and M.~Wilhelm, \emph{{On a CFT limit of planar $\gamma_i$-deformed
  $\mathcal{N}=4$ SYM theory}},
  \href{http://dx.doi.org/10.1016/j.physletb.2016.03.004}{\emph{Phys. Lett.}
  {\bf B756} (2016) 118--120}, [\href{https://arxiv.org/abs/1602.05817}{{\tt
  1602.05817}}].

\bibitem{Pomoni:2008de}
E.~Pomoni and L.~Rastelli, \emph{{Large N Field Theory and AdS Tachyons}},
  \href{http://dx.doi.org/10.1088/1126-6708/2009/04/020}{\emph{JHEP} {\bf 04}
  (2009) 020}, [\href{https://arxiv.org/abs/0805.2261}{{\tt 0805.2261}}].

\bibitem{Leigh:1995ep}
R.~G. Leigh and M.~J. Strassler, \emph{{Exactly marginal operators and duality
  in four-dimensional N=1 supersymmetric gauge theory}},
  \href{http://dx.doi.org/10.1016/0550-3213(95)00261-P}{\emph{Nucl. Phys.} {\bf
  B447} (1995) 95--136}, [\href{https://arxiv.org/abs/hep-th/9503121}{{\tt
  hep-th/9503121}}].

\bibitem{Lunin:2005jy}
O.~Lunin and J.~M. Maldacena, \emph{{Deforming field theories with U(1) x U(1)
  global symmetry and their gravity duals}},
  \href{http://dx.doi.org/10.1088/1126-6708/2005/05/033}{\emph{JHEP} {\bf 05}
  (2005) 033}, [\href{https://arxiv.org/abs/hep-th/0502086}{{\tt
  hep-th/0502086}}].

\bibitem{Frolov:2005dj}
S.~Frolov, \emph{{Lax pair for strings in Lunin-Maldacena background}},
  \href{http://dx.doi.org/10.1088/1126-6708/2005/05/069}{\emph{JHEP} {\bf 05}
  (2005) 069}, [\href{https://arxiv.org/abs/hep-th/0503201}{{\tt
  hep-th/0503201}}].

\bibitem{Beisert:2005if}
N.~Beisert and R.~Roiban, \emph{{Beauty and the twist: The Bethe ansatz for
  twisted N=4 SYM}},
  \href{http://dx.doi.org/10.1088/1126-6708/2005/08/039}{\emph{JHEP} {\bf 08}
  (2005) 039}, [\href{https://arxiv.org/abs/hep-th/0505187}{{\tt
  hep-th/0505187}}].

\bibitem{Gromov:2017cja}
N.~Gromov, V.~Kazakov, G.~Korchemsky, S.~Negro and G.~Sizov,
  \emph{{Integrability of Conformal Fishnet Theory}},
  \href{http://dx.doi.org/10.1007/JHEP01(2018)095}{\emph{JHEP} {\bf 01} (2018)
  095}, [\href{https://arxiv.org/abs/1706.04167}{{\tt 1706.04167}}].

\bibitem{Chicherin:2017frs}
D.~Chicherin, V.~Kazakov, F.~Loebbert, D.~Muller and D.-l. Zhong,
  \emph{{Yangian Symmetry for Fishnet Feynman Graphs}},
  \href{http://dx.doi.org/10.1103/PhysRevD.96.121901}{\emph{Phys. Rev.} {\bf
  D96} (2017) 121901}, [\href{https://arxiv.org/abs/1708.00007}{{\tt
  1708.00007}}].

\bibitem{Chicherin:2017cns}
D.~Chicherin, V.~Kazakov, F.~Loebbert, D.~Muller and D.-l. Zhong,
  \emph{{Yangian Symmetry for Bi-Scalar Loop Amplitudes}},
  \href{http://dx.doi.org/10.1007/JHEP05(2018)003}{\emph{JHEP} {\bf 05} (2018)
  003}, [\href{https://arxiv.org/abs/1704.01967}{{\tt 1704.01967}}].

\bibitem{Basso:2017jwq}
B.~Basso and L.~J. Dixon, \emph{{Gluing Ladder Feynman Diagrams into
  Fishnets}},
  \href{http://dx.doi.org/10.1103/PhysRevLett.119.071601}{\emph{Phys. Rev.
  Lett.} {\bf 119} (2017) 071601},
  [\href{https://arxiv.org/abs/1705.03545}{{\tt 1705.03545}}].

\bibitem{Bazhanov:2016ajm}
V.~V. Bazhanov, A.~P. Kels and S.~M. Sergeev, \emph{{Quasi-classical expansion
  of the star-triangle relation and integrable systems on quad-graphs}},
  \href{http://dx.doi.org/10.1088/1751-8113/49/46/464001}{\emph{J. Phys.} {\bf
  A49} (2016) 464001}, [\href{https://arxiv.org/abs/1602.07076}{{\tt
  1602.07076}}].

\bibitem{Klebanov:1991qa}
I.~R. Klebanov, \emph{{String theory in two-dimensions}},  in \emph{{Spring
  School on String Theory and Quantum Gravity (to be followed by Workshop)
  Trieste, Italy, April 15-23, 1991}}, pp.~30--101, 1991.
\newblock \href{https://arxiv.org/abs/hep-th/9108019}{{\tt hep-th/9108019}}.

\bibitem{DiFrancesco:1993cyw}
P.~Di~Francesco, P.~H. Ginsparg and J.~Zinn-Justin, \emph{{2-D Gravity and
  random matrices}},
  \href{http://dx.doi.org/10.1016/0370-1573(94)00084-G}{\emph{Phys. Rept.} {\bf
  254} (1995) 1--133}, [\href{https://arxiv.org/abs/hep-th/9306153}{{\tt
  hep-th/9306153}}].

\bibitem{Ginsparg:1993is}
P.~H. Ginsparg and G.~W. Moore, \emph{{Lectures on 2-D gravity and 2-D string
  theory}},  in \emph{{Proceedings, Theoretical Advanced Study Institute (TASI
  92): From Black Holes and Strings to Particles: Boulder, USA, June 1-26,
  1992}}, pp.~277--469, 1993.
\newblock \href{https://arxiv.org/abs/hep-th/9304011}{{\tt hep-th/9304011}}.

\bibitem{Nakayama:2004vk}
Y.~Nakayama, \emph{{Liouville field theory: A Decade after the revolution}},
  \href{http://dx.doi.org/10.1142/S0217751X04019500}{\emph{Int. J. Mod. Phys.}
  {\bf A19} (2004) 2771--2930},
  [\href{https://arxiv.org/abs/hep-th/0402009}{{\tt hep-th/0402009}}].

\bibitem{Ambjorn:2005wa}
J.~Ambjorn, R.~A. Janik and C.~Kristjansen, \emph{{Wrapping interactions and a
  new source of corrections to the spin-chain/string duality}},
  \href{http://dx.doi.org/10.1016/j.nuclphysb.2005.12.007}{\emph{Nucl. Phys.}
  {\bf B736} (2006) 288--301},
  [\href{https://arxiv.org/abs/hep-th/0510171}{{\tt hep-th/0510171}}].

\bibitem{Bajnok:2008bm}
Z.~Bajnok and R.~A. Janik, \emph{{Four-loop perturbative Konishi from strings
  and finite size effects for multiparticle states}},
  \href{http://dx.doi.org/10.1016/j.nuclphysb.2008.08.020}{\emph{Nucl. Phys.}
  {\bf B807} (2009) 625--650}, [\href{https://arxiv.org/abs/0807.0399}{{\tt
  0807.0399}}].

\bibitem{Janik:2010kd}
R.~A. Janik, \emph{{Review of AdS/CFT Integrability, Chapter III.5: L\'uscher
  Corrections}}, \href{http://dx.doi.org/10.1007/s11005-011-0511-z}{\emph{Lett.
  Math. Phys.} {\bf 99} (2012) 277--297},
  [\href{https://arxiv.org/abs/1012.3994}{{\tt 1012.3994}}].

\bibitem{Minahan:2002ve}
J.~A. Minahan and K.~Zarembo, \emph{{The Bethe ansatz for N=4
  superYang-Mills}},
  \href{http://dx.doi.org/10.1088/1126-6708/2003/03/013}{\emph{JHEP} {\bf 03}
  (2003) 013}, [\href{https://arxiv.org/abs/hep-th/0212208}{{\tt
  hep-th/0212208}}].

\bibitem{Bajnok:2010ke}
Z.~Bajnok, \emph{{Review of AdS/CFT Integrability, Chapter III.6: Thermodynamic
  Bethe Ansatz}},
  \href{http://dx.doi.org/10.1007/s11005-011-0512-y}{\emph{Lett. Math. Phys.}
  {\bf 99} (2012) 299--320}, [\href{https://arxiv.org/abs/1012.3995}{{\tt
  1012.3995}}].

\bibitem{Arutyunov:2014cra}
G.~Arutyunov and S.~J. van Tongeren, \emph{{$\mathrm{AdS}_5 \times
  \mathrm{S}^5$ mirror model as a string sigma model}},
  \href{http://dx.doi.org/10.1103/PhysRevLett.113.261605}{\emph{Phys. Rev.
  Lett.} {\bf 113} (2014) 261605}, [\href{https://arxiv.org/abs/1406.2304}{{\tt
  1406.2304}}].

\bibitem{Kostov:2018ckg}
I.~Kostov, D.~Serban and D.-L. Vu, \emph{{TBA and tree expansion}},  in
  \emph{{12th International Workshop on Lie Theory and Its Applications in
  Physics (LT-12) Varna, Bulgaria, June 19-25, 2017}}, 2018.
\newblock \href{https://arxiv.org/abs/1805.02591}{{\tt 1805.02591}}.

\bibitem{Ahn:2011xq}
C.~Ahn, Z.~Bajnok, D.~Bombardelli and R.~I. Nepomechie, \emph{{TBA, NLO Luscher
  correction, and double wrapping in twisted AdS/CFT}},
  \href{http://dx.doi.org/10.1007/JHEP12(2011)059}{\emph{JHEP} {\bf 12} (2011)
  059}, [\href{https://arxiv.org/abs/1108.4914}{{\tt 1108.4914}}].

\bibitem{vanTongeren:2016hhc}
S.~J. van Tongeren, \emph{{Introduction to the thermodynamic Bethe ansatz}},
  \href{https://arxiv.org/abs/1606.02951}{{\tt 1606.02951}}.

\bibitem{Gromov:2013pga}
N.~Gromov, V.~Kazakov, S.~Leurent and D.~Volin, \emph{{Quantum Spectral Curve
  for Planar $\mathcal{N} =$ Super-Yang-Mills Theory}},
  \href{http://dx.doi.org/10.1103/PhysRevLett.112.011602}{\emph{Phys. Rev.
  Lett.} {\bf 112} (2014) 011602}, [\href{https://arxiv.org/abs/1305.1939}{{\tt
  1305.1939}}].

\bibitem{Kazakov:2015efa}
V.~Kazakov, S.~Leurent and D.~Volin, \emph{{T-system on T-hook: Grassmannian
  Solution and Twisted Quantum Spectral Curve}},
  \href{http://dx.doi.org/10.1007/JHEP12(2016)044}{\emph{JHEP} {\bf 12} (2016)
  044}, [\href{https://arxiv.org/abs/1510.02100}{{\tt 1510.02100}}].

\bibitem{Zamolodchikov:1977nu}
A.~B. Zamolodchikov and A.~B. Zamolodchikov, \emph{{Relativistic Factorized S
  Matrix in Two-Dimensions Having O(N) Isotopic Symmetry}},
  \href{http://dx.doi.org/10.1016/0550-3213(78)90239-0}{\emph{Nucl. Phys.} {\bf
  B133} (1978) 525}.

\bibitem{Zamolodchikov:1978xm}
A.~B. Zamolodchikov and A.~B. Zamolodchikov, \emph{{Factorized s Matrices in
  Two-Dimensions as the Exact Solutions of Certain Relativistic Quantum Field
  Models}}, \href{http://dx.doi.org/10.1016/0003-4916(79)90391-9}{\emph{Annals
  Phys.} {\bf 120} (1979) 253--291}.

\bibitem{Fateev:1992tk}
V.~A. Fateev, E.~Onofri and A.~B. Zamolodchikov, \emph{{The Sausage model
  (integrable deformations of O(3) sigma model)}},
  \href{http://dx.doi.org/10.1016/0550-3213(93)90001-6}{\emph{Nucl. Phys.} {\bf
  B406} (1993) 521--565}.

\bibitem{Zamolodchikov:1992zr}
A.~B. Zamolodchikov and A.~B. Zamolodchikov, \emph{{Massless factorized
  scattering and sigma models with topological terms}},
  \href{http://dx.doi.org/10.1016/0550-3213(92)90136-Y}{\emph{Nucl. Phys.} {\bf
  B379} (1992) 602--623}.

\bibitem{Fendley:1993wq}
P.~Fendley, H.~Saleur and A.~B. Zamolodchikov, \emph{{Massless flows. 1. The
  Sine-Gordon and O(n) models}},
  \href{http://dx.doi.org/10.1142/S0217751X93002265}{\emph{Int. J. Mod. Phys.}
  {\bf A8} (1993) 5717--5750},
  [\href{https://arxiv.org/abs/hep-th/9304050}{{\tt hep-th/9304050}}].

\bibitem{Fendley:1993xa}
P.~Fendley, H.~Saleur and A.~B. Zamolodchikov, \emph{{Massless flows, 2. The
  Exact S matrix approach}},
  \href{http://dx.doi.org/10.1142/S0217751X93002277}{\emph{Int. J. Mod. Phys.}
  {\bf A8} (1993) 5751--5778},
  [\href{https://arxiv.org/abs/hep-th/9304051}{{\tt hep-th/9304051}}].

\bibitem{Fendley:2000bw}
P.~Fendley, \emph{{Integrable sigma models with theta = pi}},
  \href{http://dx.doi.org/10.1103/PhysRevB.63.104429}{\emph{Phys. Rev.} {\bf
  B63} (2001) 104429}, [\href{https://arxiv.org/abs/cond-mat/0008372}{{\tt
  cond-mat/0008372}}].

\bibitem{Mann:2004jr}
N.~Mann and J.~Polchinski, \emph{{Finite density states in integrable conformal
  field theories}},  \href{https://arxiv.org/abs/hep-th/0408162}{{\tt
  hep-th/0408162}}.

\bibitem{Hasenfratz:1990ab}
P.~Hasenfratz and F.~Niedermayer, \emph{{The Exact mass gap of the O(N) sigma
  model for arbitrary N is >= 3 in d = 2}},
  \href{http://dx.doi.org/10.1016/0370-2693(90)90686-Z}{\emph{Phys. Lett.} {\bf
  B245} (1990) 529--532}.

\bibitem{Bajnok:2008it}
Z.~Bajnok, J.~Balog, B.~Basso, G.~P. Korchemsky and L.~Palla, \emph{{Scaling
  function in AdS/CFT from the O(6) sigma model}},
  \href{http://dx.doi.org/10.1016/j.nuclphysb.2008.11.023}{\emph{Nucl. Phys.}
  {\bf B811} (2009) 438--462}, [\href{https://arxiv.org/abs/0809.4952}{{\tt
  0809.4952}}].

\bibitem{Polyakov:2001af}
A.~M. Polyakov, \emph{{Gauge fields and space-time}},
  \href{http://dx.doi.org/10.1142/S0217751X02013071}{\emph{Int. J. Mod. Phys.}
  {\bf A17S1} (2002) 119--136},
  [\href{https://arxiv.org/abs/hep-th/0110196}{{\tt hep-th/0110196}}].

\bibitem{Friess:2005be}
J.~J. Friess and S.~S. Gubser, \emph{{Non-linear sigma models with anti-de
  Sitter target spaces}},
  \href{http://dx.doi.org/10.1016/j.nuclphysb.2006.05.008}{\emph{Nucl. Phys.}
  {\bf B750} (2006) 111--141},
  [\href{https://arxiv.org/abs/hep-th/0512355}{{\tt hep-th/0512355}}].

\bibitem{Duncan:2007vs}
A.~Duncan, M.~Niedermaier and P.~Weisz, \emph{{Noncompact sigma-models: Large N
  expansion and thermodynamic limit}},
  \href{http://dx.doi.org/10.1016/j.nuclphysb.2007.07.020}{\emph{Nucl. Phys.}
  {\bf B791} (2008) 193--230}, [\href{https://arxiv.org/abs/0706.2929}{{\tt
  0706.2929}}].

\bibitem{Goldschmidt:1980wq}
Y.~Y. Goldschmidt and E.~Witten, \emph{{Conservation Laws in Some
  Two-dimensional Models}},
  \href{http://dx.doi.org/10.1016/0370-2693(80)91004-7}{\emph{Phys. Lett.} {\bf
  91B} (1980) 392--396}.

\bibitem{BS05}
N.~Beisert and M.~Staudacher, \emph{{Long-range psu(2,2|4) Bethe Ansatze for
  gauge theory and strings}},
  \href{http://dx.doi.org/10.1016/j.nuclphysb.2005.06.038}{\emph{Nucl. Phys.}
  {\bf B727} (2005) 1--62}, [\href{https://arxiv.org/abs/hep-th/0504190}{{\tt
  hep-th/0504190}}].

\bibitem{Balog:2001sr}
J.~Balog and A.~Hegedus, \emph{{Virial expansion and TBA in O(N) sigma
  models}}, \href{http://dx.doi.org/10.1016/S0370-2693(01)01307-7}{\emph{Phys.
  Lett.} {\bf B523} (2001) 211--220},
  [\href{https://arxiv.org/abs/hep-th/0108071}{{\tt hep-th/0108071}}].

\bibitem{Balog:2005yz}
J.~Balog and A.~Hegedus, \emph{{TBA equations for the mass gap in the O(2r)
  non-linear sigma-models}},
  \href{http://dx.doi.org/10.1016/j.nuclphysb.2005.07.032}{\emph{Nucl. Phys.}
  {\bf B725} (2005) 531--553},
  [\href{https://arxiv.org/abs/hep-th/0504186}{{\tt hep-th/0504186}}].

\bibitem{Fendley:1999gb}
P.~Fendley, \emph{{Sigma models as perturbed conformal field theories}},
  \href{http://dx.doi.org/10.1103/PhysRevLett.83.4468}{\emph{Phys. Rev. Lett.}
  {\bf 83} (1999) 4468--4471},
  [\href{https://arxiv.org/abs/hep-th/9906036}{{\tt hep-th/9906036}}].

\bibitem{Tseytlin:2003ac}
A.~A. Tseytlin, \emph{{On semiclassical approximation and spinning string
  vertex operators in AdS(5) x S**5}},
  \href{http://dx.doi.org/10.1016/S0550-3213(03)00456-5}{\emph{Nucl. Phys.}
  {\bf B664} (2003) 247--275},
  [\href{https://arxiv.org/abs/hep-th/0304139}{{\tt hep-th/0304139}}].

\bibitem{Gromov:2009tq}
N.~Gromov, \emph{{Y-system and Quasi-Classical Strings}},
  \href{http://dx.doi.org/10.1007/JHEP01(2010)112}{\emph{JHEP} {\bf 01} (2010)
  112}, [\href{https://arxiv.org/abs/0910.3608}{{\tt 0910.3608}}].

\bibitem{Zamolodchikov:1991pc}
A.~B. Zamolodchikov, \emph{{Resonance factorized scattering and roaming
  trajectories}}, \href{http://dx.doi.org/10.1088/0305-4470/39/41/S08}{\emph{J.
  Phys.} {\bf A39} (2006) 12847--12862}.

\bibitem{Zamolodchikov:2000kt}
A.~B. Zamolodchikov, \emph{{On the thermodynamic Bethe ansatz equation in
  sinh-Gordon model}},
  \href{http://dx.doi.org/10.1088/0305-4470/39/41/S09}{\emph{J. Phys.} {\bf
  A39} (2006) 12863--12887}, [\href{https://arxiv.org/abs/hep-th/0005181}{{\tt
  hep-th/0005181}}].

\bibitem{Teschner:2007ng}
J.~Teschner, \emph{{On the spectrum of the Sinh-Gordon model in finite
  volume}},
  \href{http://dx.doi.org/10.1016/j.nuclphysb.2008.01.021}{\emph{Nucl. Phys.}
  {\bf B799} (2008) 403--429},
  [\href{https://arxiv.org/abs/hep-th/0702214}{{\tt hep-th/0702214}}].

\bibitem{Zamolodchikov:1989cf}
A.~B. Zamolodchikov, \emph{{Thermodynamic Bethe Ansatz in Relativistic Models.
  Scaling Three State Potts and Lee-yang Models}},
  \href{http://dx.doi.org/10.1016/0550-3213(90)90333-9}{\emph{Nucl. Phys.} {\bf
  B342} (1990) 695--720}.

\bibitem{Klassen:1989ui}
T.~R. Klassen and E.~Melzer, \emph{{Purely Elastic Scattering Theories and
  their Ultraviolet Limits}},
  \href{http://dx.doi.org/10.1016/0550-3213(90)90643-R}{\emph{Nucl. Phys.} {\bf
  B338} (1990) 485--528}.

\bibitem{Klassen:1990dx}
T.~R. Klassen and E.~Melzer, \emph{{The Thermodynamics of purely elastic
  scattering theories and conformal perturbation theory}},
  \href{http://dx.doi.org/10.1016/0550-3213(91)90159-U}{\emph{Nucl. Phys.} {\bf
  B350} (1991) 635--689}.

\bibitem{Zamolodchikov:1991et}
A.~B. Zamolodchikov, \emph{{On the thermodynamic Bethe ansatz equations for
  reflectionless ADE scattering theories}},
  \href{http://dx.doi.org/10.1016/0370-2693(91)91737-G}{\emph{Phys. Lett.} {\bf
  B253} (1991) 391--394}.

\bibitem{Polyakov:2005ss}
A.~M. Polyakov, \emph{{Supermagnets and sigma models}},  in \emph{{Quarks,
  hadrons, and strong interactions: Gribov memorial volume. Proceedings,
  Memorial Workshop devoted to the 75th birthday of V.N. Gribov, Budapest,
  Hungary, May 22-24, 2005}}, pp.~409--428, 2005.
\newblock \href{https://arxiv.org/abs/hep-th/0512310}{{\tt hep-th/0512310}}.
\newblock \href{http://dx.doi.org/10.1142/9789812773784_0036}{DOI}.

\bibitem{Mamroud:2017uyz}
O.~Mamroud and G.~Torrents, \emph{{RG stability of integrable fishnet models}},
  \href{http://dx.doi.org/10.1007/JHEP06(2017)012}{\emph{JHEP} {\bf 06} (2017)
  012}, [\href{https://arxiv.org/abs/1703.04152}{{\tt 1703.04152}}].

\bibitem{Ikhlef:2011ay}
Y.~Ikhlef, J.~L. Jacobsen and H.~Saleur, \emph{{An Integrable spin chain for
  the SL(2,R)/U(1) black hole sigma model}},
  \href{http://dx.doi.org/10.1103/PhysRevLett.108.081601}{\emph{Phys. Rev.
  Lett.} {\bf 108} (2012) 081601}, [\href{https://arxiv.org/abs/1109.1119}{{\tt
  1109.1119}}].

\bibitem{Kazakov:2004qf}
V.~A. Kazakov, A.~Marshakov, J.~A. Minahan and K.~Zarembo,
  \emph{{Classical/quantum integrability in AdS/CFT}},
  \href{http://dx.doi.org/10.1088/1126-6708/2004/05/024}{\emph{JHEP} {\bf 05}
  (2004) 024}, [\href{https://arxiv.org/abs/hep-th/0402207}{{\tt
  hep-th/0402207}}].

\bibitem{Gromov:2006dh}
N.~Gromov, V.~Kazakov, K.~Sakai and P.~Vieira, \emph{{Strings as multi-particle
  states of quantum sigma-models}},
  \href{http://dx.doi.org/10.1016/j.nuclphysb.2006.11.018}{\emph{Nucl. Phys.}
  {\bf B764} (2007) 15--61}, [\href{https://arxiv.org/abs/hep-th/0603043}{{\tt
  hep-th/0603043}}].

\bibitem{Gross:2017aos}
D.~J. Gross and V.~Rosenhaus, \emph{{All point correlation functions in SYK}},
  \href{http://dx.doi.org/10.1007/JHEP12(2017)148}{\emph{JHEP} {\bf 12} (2017)
  148}, [\href{https://arxiv.org/abs/1710.08113}{{\tt 1710.08113}}].

\bibitem{Polchinski:1998rq}
J.~Polchinski, \emph{{String theory. Vol. 1: An introduction to the bosonic
  string}}.
\newblock Cambridge University Press, 2007.

\bibitem{Harmark:2017yrv}
T.~Harmark and M.~Wilhelm, \emph{{Hagedorn Temperature of AdS$_5$/CFT$_4$ via
  Integrability}},
  \href{http://dx.doi.org/10.1103/PhysRevLett.120.071605}{\emph{Phys. Rev.
  Lett.} {\bf 120} (2018) 071605},
  [\href{https://arxiv.org/abs/1706.03074}{{\tt 1706.03074}}].

\bibitem{Harmark:2018red}
T.~Harmark and M.~Wilhelm, \emph{{The Hagedorn temperature of AdS$_5$/CFT$_4$
  at finite coupling via the Quantum Spectral Curve}},
  \href{https://arxiv.org/abs/1803.04416}{{\tt 1803.04416}}.

\bibitem{Bajnok:2013wsa}
Z.~Bajnok, N.~Drukker, A.~Hegedus, R.~I. Nepomechie, L.~Palla, C.~Sieg et~al.,
  \emph{{The spectrum of tachyons in AdS/CFT}},
  \href{http://dx.doi.org/10.1007/JHEP03(2014)055}{\emph{JHEP} {\bf 03} (2014)
  055}, [\href{https://arxiv.org/abs/1312.3900}{{\tt 1312.3900}}].

\bibitem{Alfimov:2018cms}
M.~Alfimov, N.~Gromov and G.~Sizov, \emph{{BFKL Spectrum of N=4 SYM: non-Zero
  Conformal Spin}},  \href{https://arxiv.org/abs/1802.06908}{{\tt 1802.06908}}.

\bibitem{Cavaglia:2018lxi}
A.~Cavaglia, N.~Gromov and F.~Levkovich-Maslyuk, \emph{{Quantum Spectral Curve
  and Structure Constants in N=4 SYM: Cusps in the Ladder Limit}},
  \href{https://arxiv.org/abs/1802.04237}{{\tt 1802.04237}}.

\bibitem{Volin:2009wr}
D.~Volin, \emph{{From the mass gap in O(N) to the non-Borel-summability in O(3)
  and O(4) sigma-models}},
  \href{http://dx.doi.org/10.1103/PhysRevD.81.105008}{\emph{Phys. Rev.} {\bf
  D81} (2010) 105008}, [\href{https://arxiv.org/abs/0904.2744}{{\tt
  0904.2744}}].

\bibitem{Volin:2010cq}
D.~Volin, \emph{{Quantum integrability and functional equations: Applications
  to the spectral problem of AdS/CFT and two-dimensional sigma models}},
  \href{http://dx.doi.org/10.1088/1751-8113/44/12/124003}{\emph{J. Phys.} {\bf
  A44} (2011) 124003}, [\href{https://arxiv.org/abs/1003.4725}{{\tt
  1003.4725}}].

\bibitem{Bajnok:2008qj}
Z.~Bajnok, R.~A. Janik and T.~Lukowski, \emph{{Four loop twist two, BFKL,
  wrapping and strings}},
  \href{http://dx.doi.org/10.1016/j.nuclphysb.2009.02.005}{\emph{Nucl. Phys.}
  {\bf B816} (2009) 376--398}, [\href{https://arxiv.org/abs/0811.4448}{{\tt
  0811.4448}}].

\bibitem{Wegner}
F.~Wegner, \emph{{Anomalous Dimensions of High-Gradient Operators in the
  n-Vector Model in 2+$\epsilon$ Dimensions}}, {\emph{Z. Phys.} {\bf B78}
  (1990) 33}.

\bibitem{Balog:2003yr}
J.~Balog and A.~Hegedus, \emph{{TBA Equations for excited states in the O(3)
  and O(4) nonlinear sigma model}},
  \href{http://dx.doi.org/10.1088/0305-4470/37/5/027}{\emph{J. Phys.} {\bf A37}
  (2004) 1881--1901}, [\href{https://arxiv.org/abs/hep-th/0309009}{{\tt
  hep-th/0309009}}].

\bibitem{Gromov:2008qe}
N.~Gromov and P.~Vieira, \emph{{The all loop AdS4/CFT3 Bethe ansatz}},
  \href{http://dx.doi.org/10.1088/1126-6708/2009/01/016}{\emph{JHEP} {\bf 01}
  (2009) 016}, [\href{https://arxiv.org/abs/0807.0777}{{\tt 0807.0777}}].

\bibitem{Ahn:2008aa}
C.~Ahn and R.~I. Nepomechie, \emph{{N=6 super Chern-Simons theory S-matrix and
  all-loop Bethe ansatz equations}},
  \href{http://dx.doi.org/10.1088/1126-6708/2008/09/010}{\emph{JHEP} {\bf 09}
  (2008) 010}, [\href{https://arxiv.org/abs/0807.1924}{{\tt 0807.1924}}].

\end{thebibliography}\endgroup
\bibliographystyle{JHEP}

\end{document}